\documentclass[fleqn,usenatbib]{mnras}

\usepackage{newtxtext,newtxmath}

\usepackage[T1]{fontenc}
\usepackage{ae,aecompl}

\usepackage{graphicx}	
\usepackage{amsmath}	
\usepackage{amssymb}	
\usepackage{booktabs}

\usepackage{times}
\usepackage{verbatim}
\usepackage[dvipsnames]{xcolor}
\usepackage{amsmath}

\newcommand{\um}[1]{\ \mathrm{#1}}

\title[New radio observations
of the Cygnus Loop]{New
high-frequency
radio observations
of the Cygnus Loop supernova remnant with the Italian radio telescopes}

\author[S. Loru et al.]{
S. Loru$^{1}$\thanks{E-mail: sara.loru@inaf.it},
A. Pellizzoni$^{2}$,
E. Egron$^{2}$,
A. Ingallinera$^{1}$,
G. Morlino$^{3}$,
S. Celli$^{4,}$ $^{5}$,
\newauthor
G. Umana$^{1}$,
C. Trigilio$^{1}$,
P. Leto$^{1}$,
M.N. Iacolina$^{6}$,
S. Righini$^{7}$,
P. Reich$^{8}$,
S. Mulas$^{9}$,
\newauthor
M. Marongiu$^{2}$,
M. Pilia$^{2}$,
A. Melis$^{2}$,
R. Concu$^{2}$,
M. Bufano$^{1}$,
C. Buemi$^{1}$,
\newauthor
F. Cavallaro$^{1}$,
S. Riggi$^{1}$,
F. Schillir\`o$^{1}$
\\
$^{1}$INAF, Osservatorio Astrofisico di Catania, Via Santa Sofia 78, 95123 Catania, IT\\
$^{2}$INAF, Osservatorio Astronomico di Cagliari, Via della Scienza 5, 09047 Selargius, Italy\\
$^{3}$INAF, Osservatorio Astrofisico di Arcetri, L.go E. Fermi 5, I-50125 Firenze, Italy\\
$^{4}$Dipartimento di Fisica dell'Universit\`a La Sapienza, P.le Aldo Moro 2, 00185 Roma, Italy\\
$^{5}$INFN - Sezione di Roma, P.le Aldo Moro 2, 00185 Roma, Italy\\
$^{6}$ASI, Osservatorio Astronomico di Cagliari, Via della Scienza 5, 09047 Selargius, Italy\\
$^{7}$INAF, Istituto di Radio Astronomia di Bologna, Via P. Gobetti 101, 40129 Bologna, Italy\\
$^{8}$Max-Planck-Institut f\"ur Radioastronomie, Auf dem H\"ugel 69, 53121 Bonn, Germany\\
$^{9}$Dipartimento di Fisica, Universit\`a degli Studi di Cagliari, SP Monserrato-Sestu, KM 0.7, 09042 Monserrato, Italy\\
}
\date{Accepted XXX. Received YYY; in original form ZZZ}
\pubyear{}
\begin{document}
\label{firstpage}
\pagerange{\pageref{firstpage}--\pageref{lastpage}}
\maketitle
\begin{abstract}
Supernova remnants (SNRs) represent a powerful laboratory to study the Cosmic-Ray acceleration processes at the shocks, and their relation to the properties of the circumstellar medium.
With the aim of studying the high-frequency radio emission and investigating the energy distribution of accelerated electrons and the magnetic field conditions,  we performed single-dish observations of the large and complex Cygnus Loop SNR from 7.0 to 24.8~GHz with the Medicina and the Sardinia Radio Telescope, focusing on the northern filament (NGC~6992) and the southern shell. Both regions show a spectrum well fitted by a power-law function ($S\propto\nu^{-\alpha}$), with spectral index $\alpha=0.45\pm0.05$ for NGC~6992 and $\alpha=0.49\pm0.01$ for the southern shell and without any indication of a spectral break.
The spectra are significantly flatter than the whole Cygnus Loop spectrum ($\alpha=0.54\pm0.01$), suggesting a departure from the plain shock acceleration mechanisms, which for NGC6992 could be related to the ongoing transition towards a radiative shock.  
We model the integrated spectrum of the whole SNR considering the evolution of the maximum energy and magnetic field amplification.
Through the radio spectral parameters, we infer a magnetic field at the shock of 10~$\mu$G. This value is compatible with a pure adiabatic compression of the interstellar magnetic field, suggesting that the amplification process is currently inefficient.

\end{abstract}




\section{Introduction}
The Cygnus Loop is a bright SNR that was discovered by William Herschel in 1784. It has an integrated radio flux density of $\sim\!210\um{Jy}$ at $1\um{GHz}$ and an apparent size of $\sim\!4^{\circ}\times3^{\circ}$. 
From the Hubble Space Telescope data, the distance of this SNR was estimated by \cite{Blair_2005} as $540\um{pc}$. More recent studies, which were based on the distance of two stars located within the remnant, established the distance to the Cygnus Loop's centre at $\sim\!735\um{pc}$ \citep{Fesen_2018}.
Its age was estimated to be in the range $\sim1-2\times10^4\um{yr}$ on the basis of X-ray  (\citealt{Hester_1986}, \citealt{Levenson_1998}) and optical measurements \citep{Fesen_2018}.
Due to its large size, its location well out of the Galactic plane, and its high brightness, the Cygnus Loop is well suited for observations across the entire electromagnetic spectrum. Furthermore, this SNR presents a complex morphology, which deviates largely from its classification of typical middle-aged 
shell-type SNR (\citealt{Aschenbach_1999}, \citealt{Leahy_1997}, \citealt{Uyaniker_2004}).

In the radio band, the Cygnus Loop exhibits a large northern circular shell, which is composed of two bright partial shells and a central filament, and a bubble-like shell located in the southern part. The nature of this very peculiar morphology is still debated \citep{Uyaniker_2002}. 
\citet{Leahy_1997} derived a strongly different polarisation fraction between the bright northeastern filament (NGC 6992, $\sim2.4$~per cent) and the southern shell ($\sim39$~per cent). By exploiting also the comparison with X-ray images, they attributed the depolarisation of NGC 6992 to high-density, thermal electrons.
Spatially-resolved spectral index studies performed by \citet{Leahy_1998} revealed a flatter spectrum of the northeastern rim with respect to other Cygnus Loop features. In the same study, the authors highlighted a spectral curvature: a steepening to high frequencies in the bright filaments and a flattening in the diffuse emission regions.
More recent radio studies attributed the region-dependent variations of spectral indices and polarisation properties between the two main shells of the Cygnus Loop to their nature as two interacting SNRs \citep{Uyaniker_2002}. 
In particular, radio spectral index changes and intrinsic variations in the magnetic field configuration were observed between NGC 6992 and the southern shell, suggesting the action of different acceleration mechanisms \citep{Uyaniker_2002}. Furthermore, the X-ray and infrared emission are strong in NGC 6992, while they are very faint or completely absent in the southern shell. Also the H~$\alpha$ and O~\textsc{iii} optical lines appear brighter and more extended in the northern filament than in the southern shell, indicating different environmental conditions \citep{Uyaniker_2002}.
On the other hand, X-ray observations do not support the two SNRs scenario (\citealt{Katsuda_2011}, \citealt{Leahy_2013}), highlighting how the X-ray emission shows a smoothly-gradual change between the northern and the southern shell, which was interpreted as the evidence of an asymmetry in the whole Cygnus Loop structure \citep{Aschenbach_1999}.
Furthermore, in a recent multi-wavelength investigation, \cite{Fesen_2018} conclude that there is 
no  morphological evidence for two separated SNRs 
in the Cygnus Loop. In particular, they rely on the lack of X-ray emission at the putative interacting interface region between the two shells, and on the morphological connection between the main optical filaments of the southern shell and those of the northern part. All these characteristics make the physical interpretation of the two regions quite controversial.
This challenging morphology makes the Cygnus Loop an ideal laboratory to study SNR shock conditions arising from different interacting structures in the interstellar medium (ISM; \citealt{Hester_1994}, \citealt{Leahy_2002}, \citealt{Leahy_2004}). 
Indeed, the investigation of the correlation between flux density and spectral radio slope in the specific SNR macro-regions is useful to reach a complete understanding of complex objects like SNRs. 
This allows us to disentangle
possible magnetic enhancement processes from spectral variation due to the energy distribution of the synchrotron emitting particles or to other emission mechanisms that could become significant at high radio frequencies.

Although the Cygnus Loop represents an interesting scientific case, high-resolution radio flux density measurements are available only up to $\sim\!5\um{GHz}$, because of the technical difficulties in performing radio continuum observations of such a large source. \cite{Sun_2006} derived an integrated spectral index of $0.40\pm0.06$ between 0.408 and $4.8\um{GHz}$, and ruled out any possible global spectral steepening within this frequency range.
However, a steepening in the integrated spectrum, like the one observed for the similar SNR S~147 \citep{Fuerst&Reich_1986}, could be expected.
Spatially-resolved spectral index studies revealed, for both the SNRs, the existence of a spectral variation between filamentary and diffuse structures, which was ascribed to the compression of the Galactic magnetic field across the filaments. In the case of S~147, the break observed in the integrated spectrum around 1~GHz was interpreted as a consequence of the spatial differences in the
compressed magnetic field \citep{Uyaniker_2004}.
It is therefore of interest to firmly establish whether there is a spectral break for the Cygnus Loop to better constrain the particle maximum energy and the magnetic field, and compare them with those of SNRs of the same age class.
Sensitive maps above $\sim\!5\um{GHz}$ could be crucial in investigate this aspect.
On the basis of this scientific interest, we performed single-dish observations of the Cygnus Loop in order to obtain sensitive images of this SNR up to high-radio frequencies.

Here, we present the observations performed on the whole Cygnus Loop with the Medicina radio telescope\footnote{\hyperlink{http://www.med.ira.inaf.it/}{http://www.med.ira.inaf.it/}} at 8.5 GHz, providing the highest-frequency map  of this SNR ever obtained with a single-dish radio telescope.
We use the Cygnus Loop radio data available in the literature, including our measurement at 8.5 GHz, and the $\gamma$-ray data presented in \cite{Katagiri_2011} to model the emission spectrum of the particles accelerated in this SNR.
Indeed, the combined study of radio and $\gamma$-ray spectra allows us to investigate the cosmic-ray (CR) acceleration mechanisms, taking into account both of the leptonic and the hadronic contributions.
We adopt the model developed by \cite{Celli+2019} and \cite{Morlino-Celli2020} to constrain the maximum energy of the particles accelerated at the remnant shock and the evolution of magnetic field strength.
We also present the observations carried out at 7.0, 18.7 and 24.8~GHz with the Sardinia Radio Telescope\footnote{\hyperlink{http://www.srt.inaf.it/}{http://www.srt.inaf.it/}} (SRT) of two 
selected regions of the Cygnus Loop: the northern-bright filament and the southern shell. 
Our observations allow us to investigate the energetics of the accelerated particles and on the possible emission mechanisms that might compete with the synchrotron emission to produce the radio
continuum emission at these frequencies.

In Sect.~\ref{Sec: Observations}, we describe the observations carried out with the Medicina and SRT telescopes and the main steps of the data reduction and analysis. The results in terms of final calibrated images and flux density measurements are presented in Sect.~\ref{Section:Results}.  In Sect.~\ref{Sec: Cyg Loop spectral analysis}, we discuss the spectral analysis performed on the whole Cygnus Loop. Sect.~\ref{Sec: Spectral analysis of NGC 6992 and the southern shell} is dedicated to the spectral analysis of the regions NGC 6992 and the southern shell.
In Sect.~\ref{sec:model}, we describe the model used to investigate the Cygnus Loop's non-thermal emission and related results.
We summarise our conclusions in Sect.~\ref{Sec: conclusion}.

\section{Observations and data analysis}
\label{Sec: Observations}

\begin{table*}
\noindent
	\centering
	\caption{Summary of the observations of the Cygnus Loop carried out with the Medicina radio
telescope and SRT. `N.maps', `Obs. time', `Freq.', `BW', `HPBW' and `Map size' indicate the number of maps, the total observation time (including overheads), the central frequency, the bandwidth, the half power
beam width and the size of the maps in RA and Dec directions, respectively. A single map is intended as a complete scan (RA or Dec) on the source.}
	\label{tab:obs_table}
	\begin{tabular}{|l|c|c|c|c|c|c|c|c|c|c|} 
		\hline
		\hline
	 Radio & Observing & Target & N.maps & Obs. time & Freq. & BW & HPBW & Map size & Map size \\
	     telescope & date &  & & (h)  &(GHz) & (MHz) & (arcmin)&  (RA) ($^{\circ}$,$^{\circ}$) &  (Dec) ($^{\circ}$,$^{\circ}$)   \\
\hline
\hline
Medicina & 2017 June & Cygnus Loop & 16 & 25.6 & 8.5 &  680 & 4.77 & 5$\times$6  & 5$\times$6 \\ 
   &  19, 21, 24, 26 &   &  &  &  & &  & & \\ 
  & 2017 Aug & Cygnus Loop &16 & 25.6 & 8.5 &  250 & 4.77 &5$\times$6 & 5$\times$6  \\ 
  &  28, 29, 30, 31 &  &  &   &  &  &  \\ 

 \hline
    SRT &  2019 Jan 24, Mar 21 & NGC 6992 & 6 & 9.9 & 7.0 &  1400 & 2.71 & 1.9$\times$ 1.9 & 1.3$\times$2.5\\
    & 2018 Dec 28 &  NGC 6992 & 1 & 1.7& 18.7 &  1400 & 0.99 & 1.9$\times$ 1.9 & -\\
    & 2018 Feb 23 &  NGC 6992 & 1 & 1.8 & 24.8 &  1400 & 0.75 & 1.9$\times$ 1.9 & -\\
    & 2019 Jan 31, May 02, Oct 22-24 & southern shell & 12 & 23.4 & 7.0 &  1400 & 2.71 & 2.2$\times$ 1.8 & 1.6$\times$2.5\\
    & 2019 Feb 05 & southern shell & 3 & 5.6 & 24.8 &  1400 & 0.75 & 2.2$\times$ 1.8 & 1.6$\times$2.5\\

 \hline
 \hline
\end{tabular}
\end{table*}

 \begin{table}
\noindent
	\centering
	\caption{Flux densities of the calibrators at the Medicina and SRT observing frequencies obtained by interpolating the values proposed by \citet{Perley_2013}.}
	\label{tab:Calibrators}
	\begin{tabular}{|c|cccc|} 
		\hline
		\hline
Calibrator	 & 7.0 GHz  &  8.5 GHz & 18.7 GHz & 24.8 GHz \\

  \\
 \hline
 \hline
 
3C286 & 5.8 Jy  & 5.0 Jy & 2.9 Jy & 2.3 Jy\\
3C295 & 4.2 Jy  & 3.3 Jy & 1.2 Jy & 0.8 Jy \\
3C147 &  5.4 Jy & 4.4 Jy & 2.1 Jy & 1.6 Jy\\
3C48 &  3.8 Jy & 3.1 Jy & 1.5 Jy & 1.1 Jy\\
3C123 &  11.2 Jy & 9.2 Jy & 4.0 Jy & 2.9 Jy\\
NGC 7027 & 5.6 Jy  & 5.8 Jy & 5.5 Jy & 5.4 Jy\\

\hline 
\hline
	\end{tabular}
\end{table}
\subsection{Medicina observations}

We observed the Cygnus Loop with the 32-m Medicina radio telescope between June and August 2017\footnote{Program code Medicina 14-17, PI: S.Loru.}. The observations were performed at the central frequency of 8.5~GHz ($X$-band) using the total-power continuum backend. 
The bandwidth was 680 MHz during the observing sessions of June, but it was subsequently reduced to 250 MHz due to the observed radio frequency interference (RFI).
The observations are summarised in Table \ref{tab:obs_table}.

The `On-the-fly' (OTF) observing technique was used to map a $5^{\circ}\times6^{\circ}$ area along RA-Dec scan directions. The map size was chosen to be twice the source extension, to properly identify and subtract the background baseline component. We set the scanning velocity to 4~$\mathrm{arcmin\ sec^{-1}}$ and sampling interval to 20~ms.
Two consecutive scans were separated by an interleave of 4.8~arcmin, which implies one passage per beam size at that observing frequency.
We chose this parameter to minimise the long time necessary to complete a map on such a wide source, and to rely on system and weather stability during the observation.
On the other hand, oversampling with respect to the beam size is required to perform a
direct evaluation of statistical errors, and to properly reject the corrupted data affected by RFI. This
leads to the achievement of improved accuracy in the final images, as demonstrated by our previous experience in single-dish imaging of Galactic sources (\citealt{Egron_2016}, \citealt{Egron_2017}, \citealt{Loru_2018}).
For this reason, we performed four complete maps for each observing session, which allowed us to acquire $\sim$240 samples per beam once merged.
The offset between the four maps acquired in each observing session is of $1.2$~arcmin and assures a proper Nyquist sampling.
The aforementioned parameters implied a total duration of a target observation (meant as a RA+Dec map) of about 3.2 hours, including slew and dead time.
We observed the point-like flux density calibrators (3C286, 3C295, 3C147, 3C48, 3C123 and NGC~7027) through repeated cross-scans at the beginning and at the end of each observing session.

\subsection{SRT observations}

The observations were carried out in the framework of our SRT observing program\footnote{Program code SRT 22-18, PI: A.Pellizzoni.} of the 2018B semester, which was focused on the high-frequency investigation of a wider sample of middle-aged and young SNRs and the precise modelling of the region-dependent continuum spectral indices.
SRT observations of the Cygnus Loop were carried out between December 2018 and October 2019 at the central frequencies of 7.0 GHz ($C$-band), 18.7 GHz and 24.8 GHz ($K$-band). The data were recorded by using the spectro-polarimetric backend SARDARA (SArdinia Roach2-based Digital Architecture for Radio Astronomy, \citealt{Melis_2018}) in full-Stokes mode with 1024 spectral channels and the maximum bandwidth available (1.4 GHz both for $C$- and $K$-band), which maximises the signal-to-noise ratio and increases the spectral coverage. A summary of the observing sessions and parameters with SRT are given in Table \ref{tab:obs_table}.
We observed the target at elevations $>\!20^{\circ}$ to avoid significantly pointing errors and beam shape instability effects. 

The SRT $K$-band receiver is characterised by a seven-feed system. 
The receiver was used in the Best Space Coverage configuration \citep{Bolli_2015} that automatically rotates the dewar to optimally cover the scanned area (scan spacing $\sim$0.87~arcmin, comparable to the half power beam width, HPBW). The observational strategy and data analysis procedure that is required to observe extended sources with this configuration is described in \citet{Loru_2018}.

\begin{figure}
    \centering
    \includegraphics[width=9.3cm]{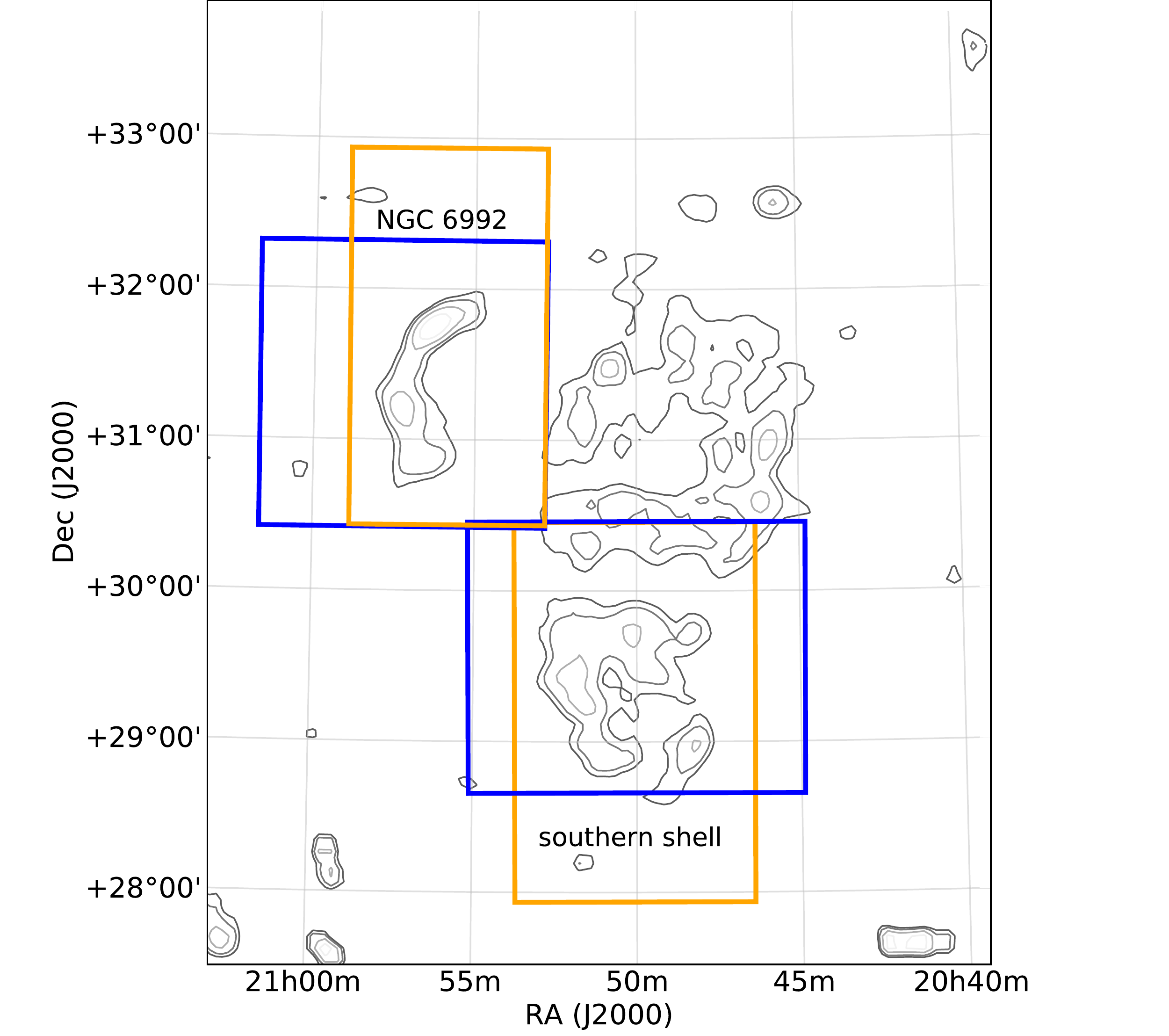}
    \caption{Representation of the regions of the Cygnus Loop (NGC 6992 and southern shell) that we observed with SRT at 7.0, 18.7  and 24.8~GHz. The blue and orange boxes indicate the maps in RA and Dec directions, respectively. The grey contours mark the continuum emission detected with Medicina at 8.5~GHz, which correspond to the intensity levels of $75$, $115$ and $189$~mJy~beam$^{-1}$.}
    \label{fig:observing_regions}
\end{figure}

Due to visibility constraints, it is not feasible to map the entire Cygnus Loop at high frequencies ($\gtrsim$7.0 GHz) in a single long observing session. Atmospheric variations might have a strong impact on the image quality in case of very long observing duration.
For this reason, we
decided to separately observe two interesting regions of the Cygnus Loop: the northern-bright filament (NGC 6992) and the southern shell. 
We adopted the well-established OTF mapping strategy in use for the INAF single-dish radio telescope network (\citealt{Egron_2017}, \citealt{Loru_2018}).
We decided to perform RA and Dec maps with different size, which are indicated respectively by the blue and orange boxes in Fig.~\ref{fig:observing_regions}, to ensure that about 50~per cent of the
scan's length/duration is free from significant source contribution. This makes the baseline subtraction easier and, at the same time, minimises the observing time. 
For all the observing frequencies, we set a scanning velocity and a sampling interval of $5\um{arcmin\ s}^{-1}$ and 20~ms, respectively. 
Two consecutive 7.0-GHz scans were separated by an interleave of 0.6~arcmin, which implies four passages per beam size, and about 27 samples beam$^{-1}$ scan$^{-1}$ were recorded. 
The same offset was adopted for the observations at 18.7 and 24.8 GHz, implying in this case one passage per beam for each feed of the receiver. We made this choice in order to minimise the observing time, considering that fast mapping mitigates the effects of time-based atmospheric opacity variations on the image quality, which especially affect the high-frequency observations \citep{Navarrini_2016}.
On the other hand, the seven-feed $K$-band receiver
configuration significantly increased the amount of data acquired for each map, and allowed us to obtain about $140\um{sample\ beam}^{-1}$ (if we consider the proper operation/contribution of all 7 feeds), a proper Nyquist sampling and an exposure time of $5.3\um{s\ beam}^{-1}$ for a full RA or Dec map. 
The $K$-band observations were performed on NGC 6992 at the two central frequencies of 18.7 and 24.8 GHz, while the southern shell was observed only at 24.8 GHz due to the strong RFI found at the lower $K$-band frequency after December 2018. 
Despite SRT observations were performed in `shared-risk mode' (for which observations under optimal weather conditions are not guaranteed), all the data presented here were acquired in the recommended opacity conditions ($\tau$<0.1 neper), which are necessary to guarantee the good quality of the $K$-band observations.

\subsection{Data reduction and analysis}
\label{Section:Data reduction and analysis}

We performed the data reduction with the SRT Single-Dish Imager (SDI) software, written in IDL and suitable for all Medicina and SRT receivers/backends (\citealt{Prandoni_2017}, \citealt{Egron_2016}, \citealt{Marongiu_2020_methods}).
This tool provides an automatic pipeline (quicklook analysis) and interactive tools for data inspection, baseline removal, RFI rejection and image
calibration (standard analysis). 
At the end of these procedures, SDI produces standard FITS images.
All final images were produced in units of mJy~beam$^{-1}$, and using the cubehelix colour scheme \citep{Green_2011}.

In the case of the Medicina observations, which were carried out with the total-intensity backend, we adopted the standard data reduction procedure as described in detail in \citet{Egron_2017}.
We performed cross-scan observations on the calibrators and used the flux density measurements and the polynomial expressions proposed by \cite{Perley_2013} to reconstruct/extrapolate their flux density at the observed frequency (see Table \ref{tab:Calibrators}).
In order to guarantee the consistency between calibrators and target in terms of the gain stability and observing condition, we chose the calibration factors by requiring that: i) they have same backend attenuation
parameters; ii) their observations are performed within 12 h (or less in case of changing weather) from each target scan epoch.
We noted the presence of persistent RFI that affected the 8.5-GHz band. 
The decision to reduce the bandwidth from 680 MHz to 250 MHz allowed us to mitigate the problem.  
The data related to the left polarisation channel were found to be strongly noisy, and we discarded them in order to avoid their negative impact on the final image quality.

The SRT observations at 7.0 and 18.7-24.8 GHz used the spectral-polarimetric backend SARDARA. This allowed us to complement the `spatial' RFI rejection procedure (as described in \citealt{Egron_2017}) with a `spectral' RFI rejection. A specific SDI routine is dedicated to the automated search for
outliers in each scan-sample's spectrum, which are dynamically
identified as RFI and rejected. After this procedure, the
data are averaged into a single continuum channel, and they can be processed with the same data reduction procedure described for the total-intensity Medicina data.
Despite the two different RFI removal processes, the not always ideal weather conditions introduced a background noise contribution that could not be deleted in the data reduction phase, especially in the case of a weak target signal. We have accounted for these effects with the background subtraction procedure described in Sect. \ref{Section:Results}.

The multi-feed data at 18.7 and 24.8~GHz were calibrated considering each feed characterised by a specific efficiency (\citealt{Orfei_2010,Loru_2018}). 
For this purpose,
we used calibrator maps of each SRT observing session in order to calculate the ratios between the expected flux of the calibrator and the peak counts related to each feed. We then applied these scaling factors to the related single feed maps of the target.

The uncertainties associated with our flux density measurements include the statistical errors and the errors related to the calibration procedure, which are added in quadrature. We calculated the first term as the product of the standard deviation
associated with the region used to estimate the background contribution 
and the square root of the number of beam solid angles that are contained in the extraction area of the target. The calibration errors were estimated as the standard deviation on the calibration factors. These are strongly related to the gain stability of the receiver and the observing conditions that characterised each session. We calculated a calibration error of 4~per cent for the 7.0-GHz data, 6~per cent for the data at 18.7 GHz and of 14~per cent for those at 24.8 GHz.

{}

\section{Results}
\label{Section:Results}

\begin{figure}

    \includegraphics[width=\columnwidth]{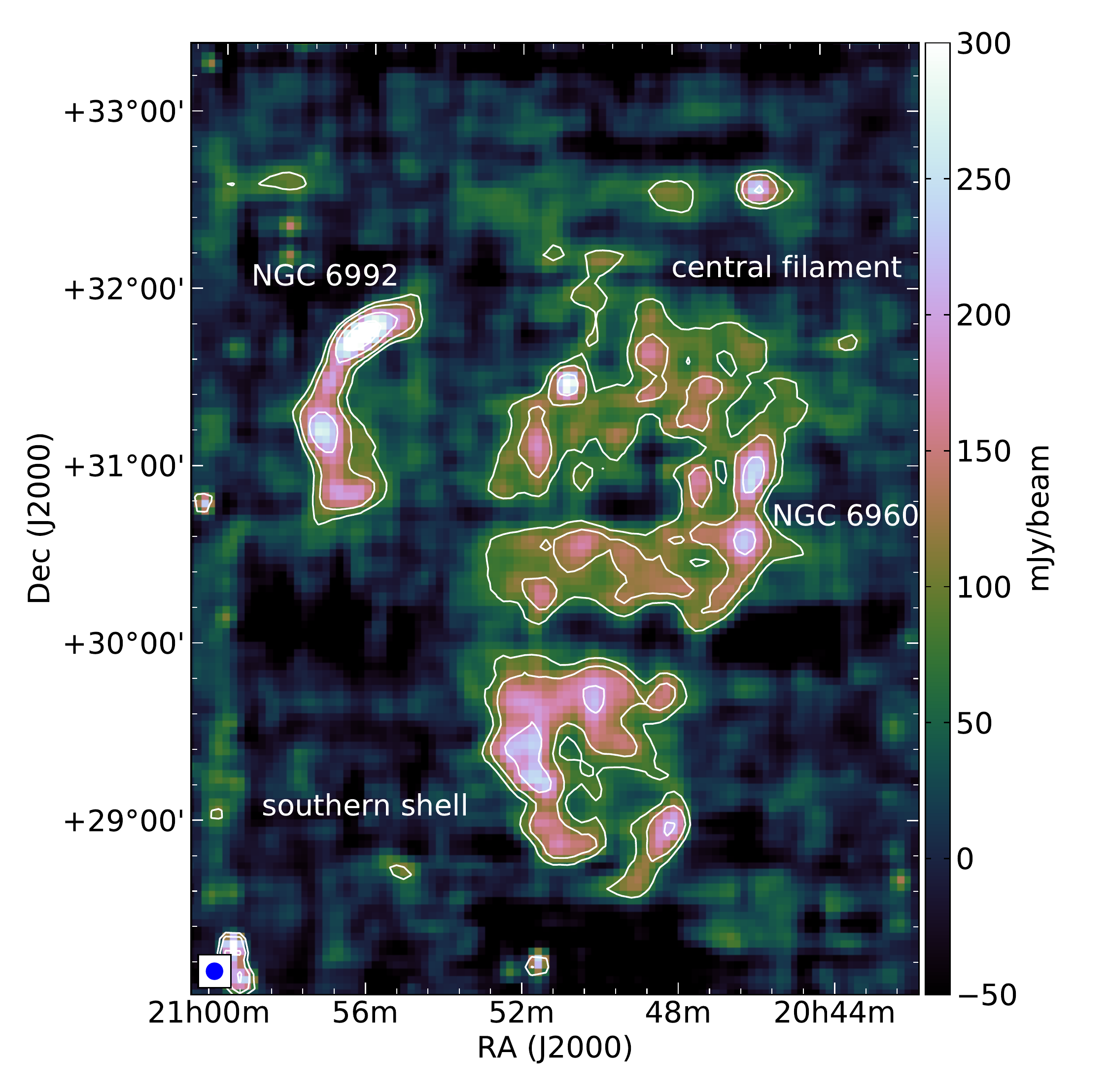}
    \caption{Map of the Cygnus Loop SNR obtained with the Medicina radio telescope at 8.5 GHz. The image was produced with pixel size of 2.4~arcmin (about 1/2 of the HPBW), to which we applied a gaussian smooth resulting in a final map resolution of 5.34~arcmin. The smoothed beam size is indicated by the blue circle on the bottom-left corner.
    The white contours highlight the major features towards the Cygnus Loop, corresponding to the intensity levels of $75$, $115$ and $189$~mJy~beam$^{-1}$.}
    \label{Cyg_Med}
\end{figure}

The map of the whole Cygnus Loop SNR at 8.5 GHz is shown in Fig.~\ref{Cyg_Med}.
It was obtained by merging and averaging all the maps of the Medicina observing sessions reported in Table \ref{tab:obs_table}.
This is the highest-frequency map of the entire Cygnus Loop SNR obtained so far with a single-dish telescope.
The Cygnus Loop results well detected, and we can distinguish the two prominent shells (NGC~6960 and NGC~6992) and the central filament, which constitutes the northern remnant, as well as the southern remnant.
These regions are highlighted by white contours in Fig.~\ref{Cyg_Med}.

Our data analysis procedure ensures a zero-mean flux associated with the regions devoid of unrelated source contamination. Despite this, we carefully chose the target extraction region in order to discard possible contributions from near-unrelated sources and RFI features affecting the image. We considered a polygonal region encompassing the SNR emission in order to carry out the integrated flux density measurement. We obtained a continuum flux density of 54$\pm$4~Jy and an image rms of $24\um{mJy\ beam}^{-1}$ (calculated in a map area free from source contribution and RFI contamination).
We also exploited our 8.5-GHz map to derive the flux densities related to the two region of interest. We calculated a flux density of 7.5$\pm$0.9~Jy for NGC~6992 and 16.6$\pm$1.5~Jy for the southern shell. 

\begin{figure*}
  \includegraphics[width=6.0cm]{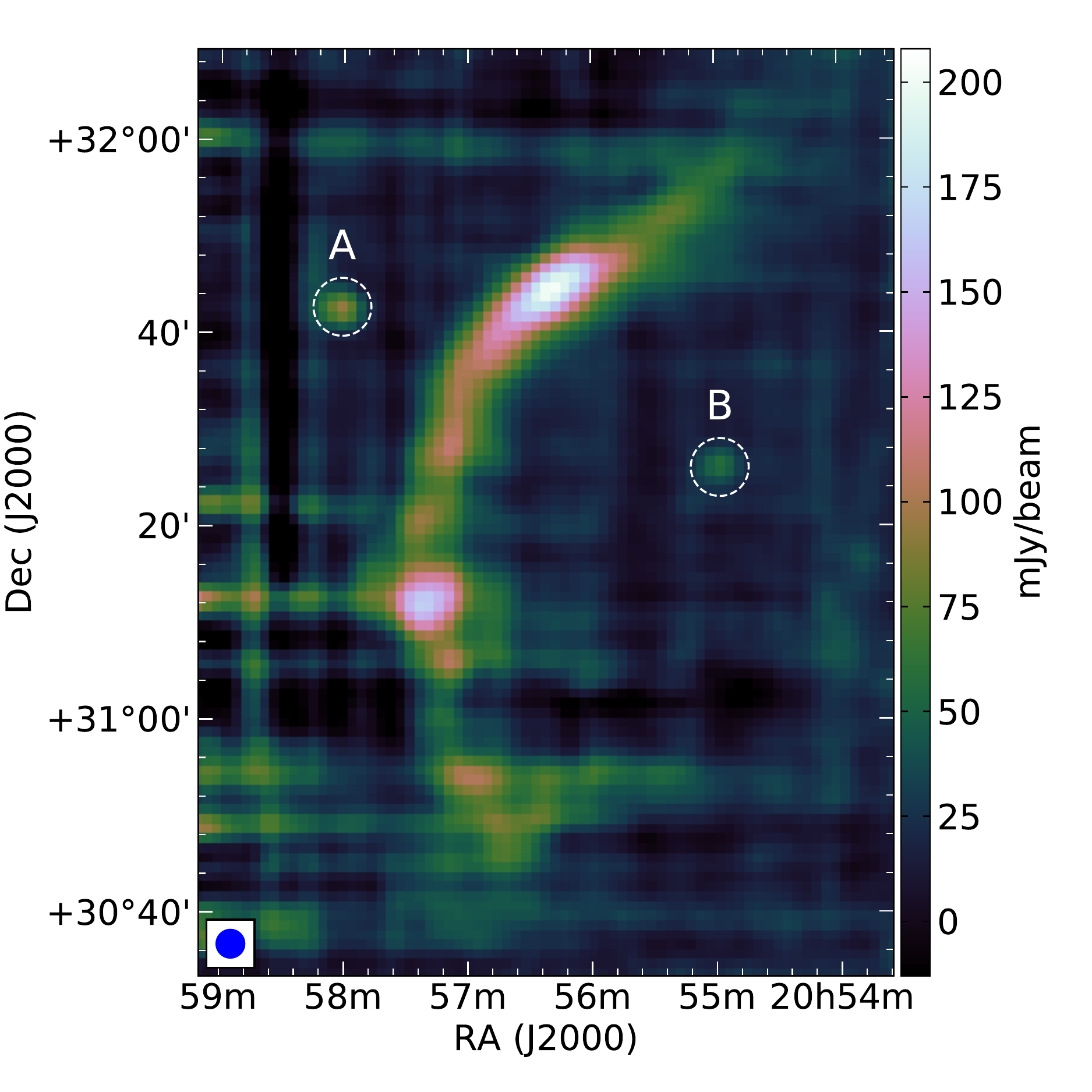}
  \hspace{-0.2cm}
  \includegraphics[width=6.0cm]{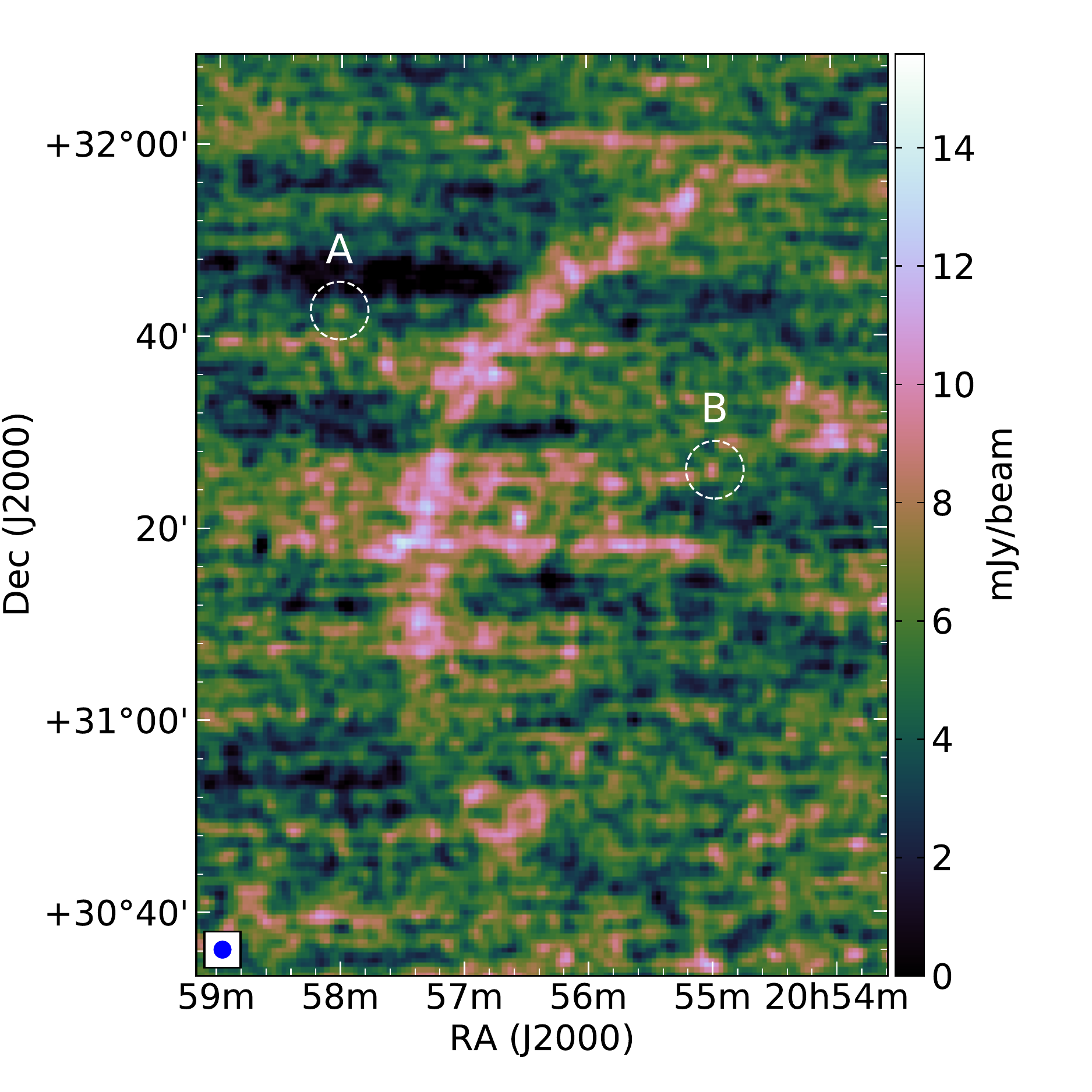}
  \hspace{-0.4cm}
  \includegraphics[width=6.0
  cm]{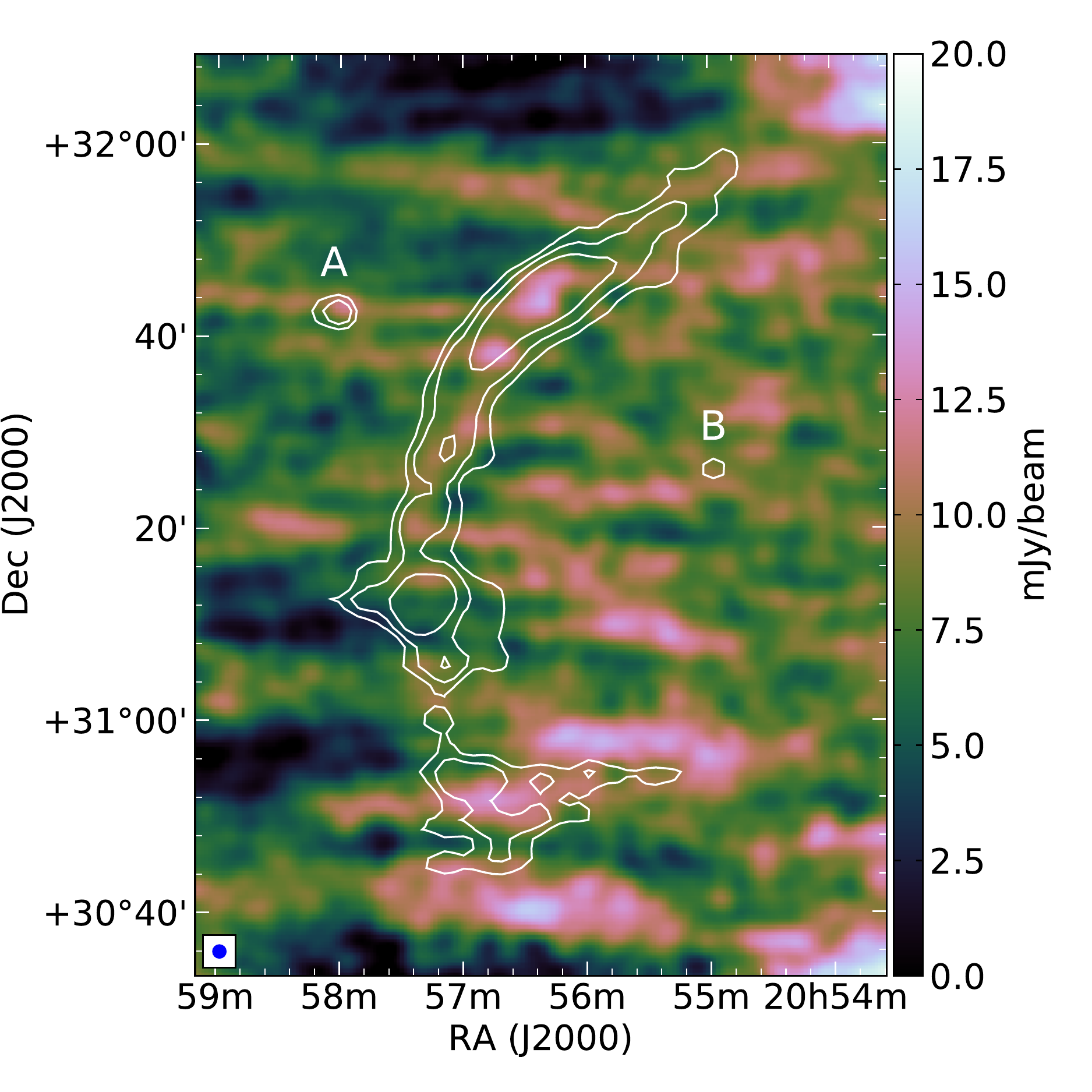}
\caption{Maps of of the bright filament NGC 6992 obtained with SRT at 7.0 GHz (\textit{left}), 18.7 GHz (\textit{middle}) and 24.8 GHz (\textit{right}). 
The maps were produced with a beam size of 2.71, 0.99 and 0.75~arcmin, and a pixel size of 1.0, 0.45 and 0.35~arcmin at 7.0, 18.7 and 24.8~GHz, respectively. 
We applied a gaussian smooth resulting in a final map resolution of 2.89~arcmin at 7.0~GHz, 1.67~arcmin at 18.7~GHz and 1.29~arcmin at 24.8~GHz. The smoothed beam is indicated by the blue circle on the bottom-left corner.
The dashed circles indicate the point sources NVSS J205800+314231 (A) and NVSS J205458+312614 (B), respectively. The white contours in the 24.8-GHz image correspond to the 7.0-GHz intensity levels of 36~mJy~beam$^{-1}$, 72~mJy~beam$^{-1}$ and 108~mJy~beam$^{-1}$.
}\label{CygB_maps}
 \end{figure*}

\begin{figure*}
  \includegraphics[width=\columnwidth]{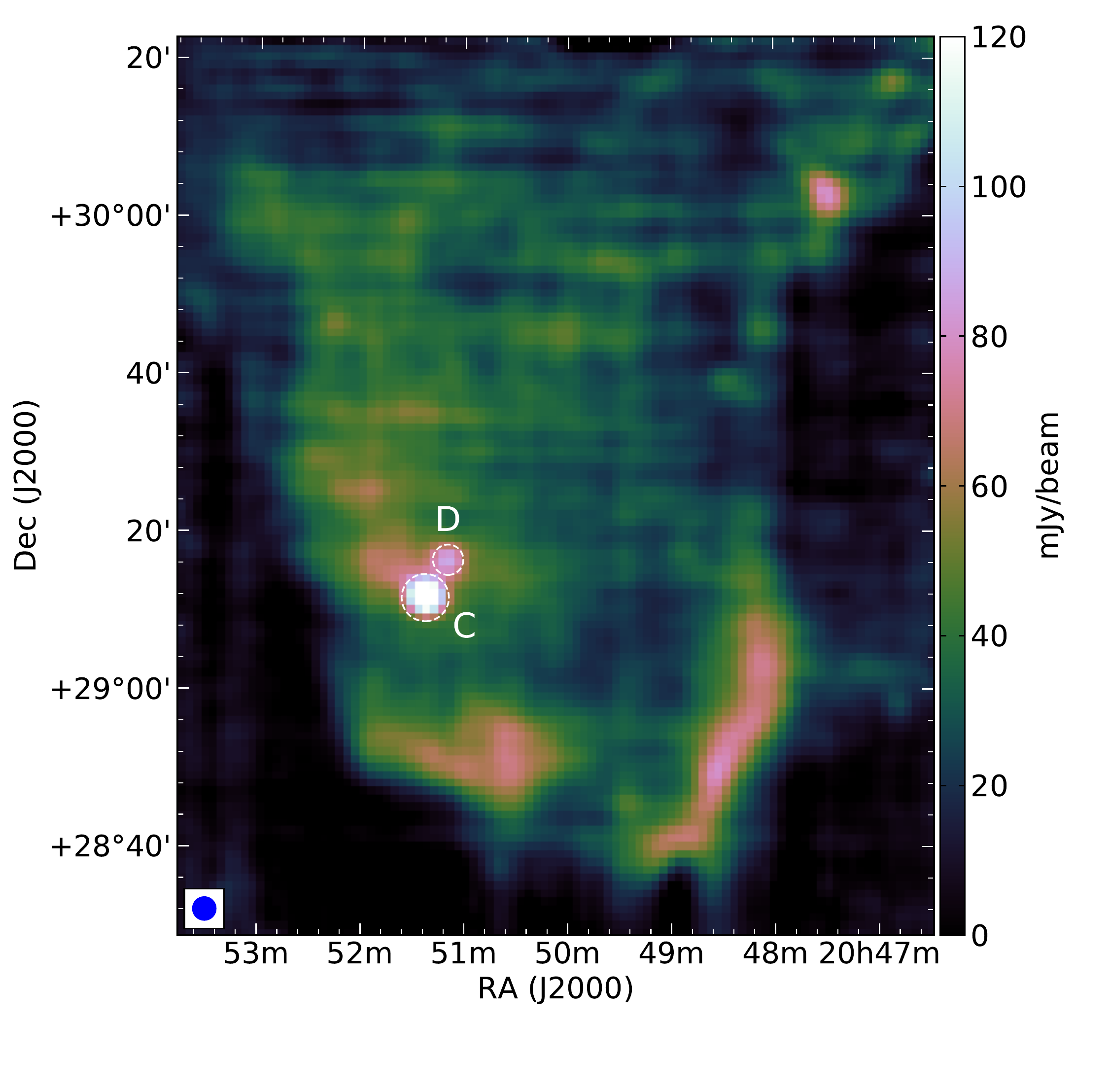}
  \hspace{-0.4cm}
  \includegraphics[width=\columnwidth]{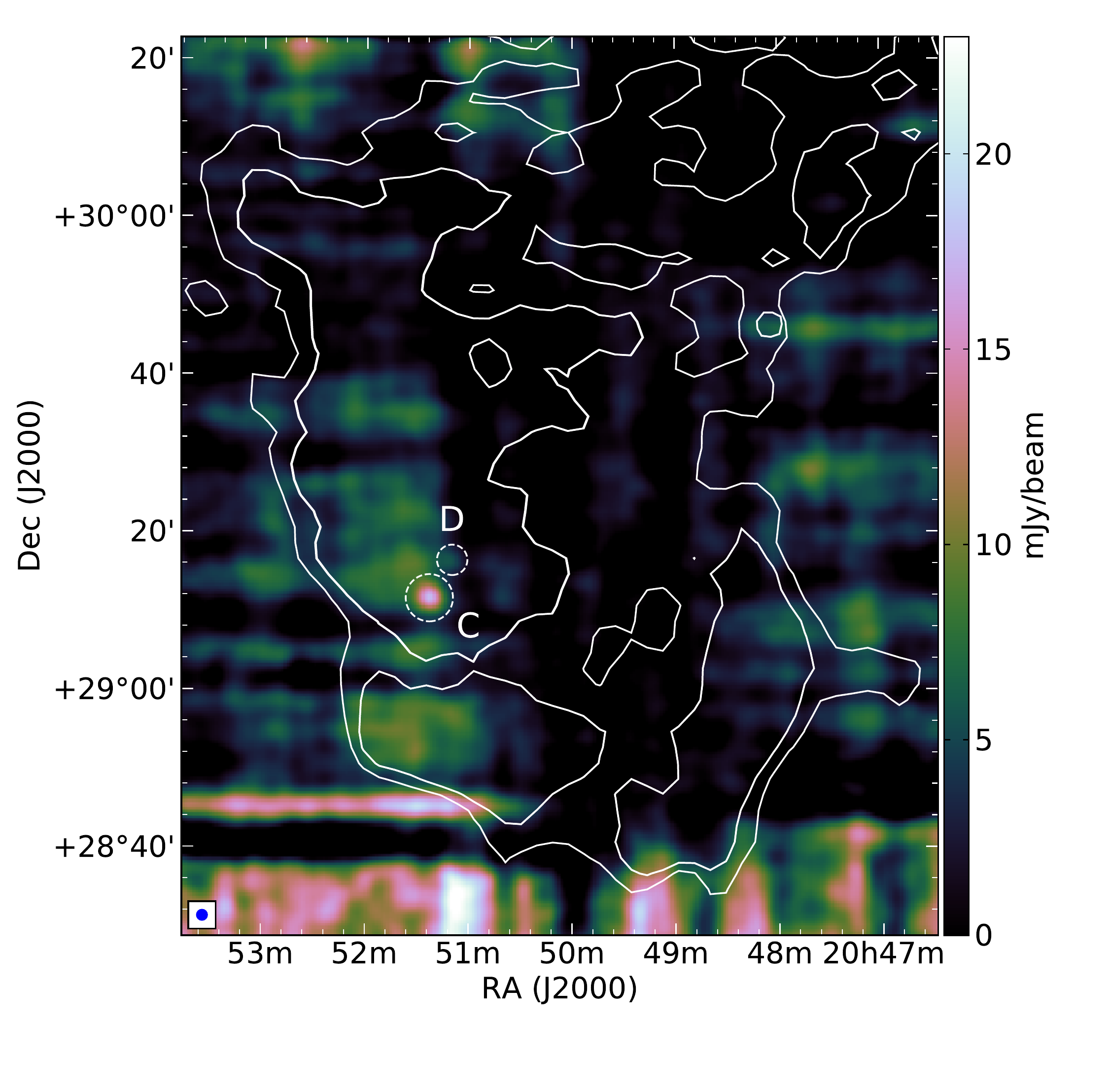}
\caption{Maps of of the southern shell obtained with SRT at 7.0 GHz (\textit{left}) and 24.8 GHz (\textit{right}). 
The maps were produced with a beam size of 2.71 and 0.75~arcmin, and a pixel size of 1.0 and 0.35~arcmin at 7.0 and 24.8~GHz, respectively. We applied a gaussian smooth resulting in a final map resolution of 2.89~arcmin at 7.0~GHz and 1.29~arcmin at 24.8~GHz. 
The smoothed beam is indicated by the blue circle on the bottom-left corner.
The dashed circles indicate the point sources NVSS J205122+291140 (C) and NVSS J205109+291629 (D), respectively. The white contours in the 24.8-GHz image correspond to the 7.0-GHz intensity levels of 21~mJy~beam$^{-1}$ and 34~mJy~beam$^{-1}$.}\label{CygA_maps}
 \end{figure*}

The SRT images of the NGC~6992 region at 7.0, 18.4 and 24.8~GHz, and of the southern shell at 7.0 and 24.8~GHz are shown in Fig.~\ref{CygB_maps} and Fig.~\ref{CygA_maps}, respectively. These are the first high-resolution images of the two regions at these frequencies.
The map resolution is equal to 2.71, 0.99 and 0.75~arcmin at 7.0, 18.7 and 24.8~GHz, respectively.
We obtained these images by averaging, for each frequency, the maps of the different observing sessions. 
The data sets selected to obtain
the final images are summarised in Table \ref{tab:obs_table}. 
Regarding the SRT observations, the total number of maps for each observing session was at last determined by the impact of the weather conditions. Especially in the case of  observations at 18.7 and 24.8 GHz, the temporal variations of the atmospheric opacity strongly affect the image quality during each observing session. Hence, we discarded the maps that contribute negatively to the final image accuracy and rms. 

We used the SRT data sets to further investigate the morphology and the spectra of NGC 6992 and the southern shell. 
The maps of NGC 6992 at 7.0 GHz, 18.7 GHz and 24.8 GHz are compared in Fig.~\ref{CygB_maps}. 
The 7.0-GHz map of NGC 6992 (Fig.~\ref{CygB_maps}, \textit{left}) reveals the clumpy emission of the southern part of the filament, probably ascribed to the interaction of the shock wave with smaller discrete clouds \citep{Fesen_2018}.
The coincident indented structures observed in the X-ray maps are indicative of a blast wave significantly hampered by dense clumps of gas, which are photoionized by the shock precursor \citep{Levenson_1998}. The northern part is instead more uniform and bright, resulting from the interaction between the remnant and a more extended cloud \citep{Levenson_1999}.
The NGC 6992 radio emission is significantly weaker at the higher frequencies. Due to the smaller beam size, in both 18.7- and 24.8-GHz images (Fig.~\ref{CygB_maps}, \textit{middle} and \textit{right}) the filamentary structure appears thinner, and the highly noisy background makes it difficult to distinguish the morphological features. 
We overlaid 7.0-GHz contours to the 24.8-GHz map (Fig.~\ref{CygB_maps}, \textit{right}) to make the visual identification of the source easier, and distinguish it from the strong and unrelated noise and RFI contribution.

Despite this, the northern filament is detected at 7.0, 18.7 and 24.8 GHz, but the attenuation of the astronomical signal, due to sky conditions (humidity level) not always ideal, and the weakening of the SNR emission with increasing frequency, make the precise baseline subtraction difficult. 
This implies that the baseline-subtracted pixels in the regions free from source contamination have non-zero mean flux density. In this case, we applied the background subtraction to obtain correct flux density measurements. In order to estimate the background contribution, we considered the flux density of a region surrounding the filament,  excluding the other regions of the Cygnus Loop, and multiplied it by the ratio between the extraction area of the source and the background area.
We calculated the flux density of NGC 6992 at 7.0, 18.7 and 24.8 GHz by considering the same extraction region used for the 8.5-GHz map. In this way, we estimated a flux density of 8.7$\pm$0.6~Jy, 5.6$\pm$0.7~Jy and 4.5$\pm$0.9~Jy at 7.0, 18.7 and 24.8~GHz, respectively.

The SRT 7.0-GHz image of the southern shell (Fig.~\ref{CygA_maps}, \textit{left}) revealed a non-uniform structure. We detected three bright regions: the western boundary and the two emission knots in the eastern side. The SNR shock appears discontinuous between these regions. The eastern feature is most likely attributed to a fully radiative shock \citep{Levenson_1999}.
The structure is difficult to be detect in the 24.8-GHz image, because of its lower signal-to-noise ratio. It is indicated by the white contours in Fig.~\ref{CygA_maps} (\textit{right}). 
We point out that the southern shell region includes the shell-like structure, clearly visible in the radio band, and the north-western structure observed both in radio and X-ray emission \citep{Aschenbach_1999}.
We calculated the 24.8-GHz flux density of the southern shell by considering the same extraction region adopted for the maps at 7.0 and 8.5~GHz. In the same way as described above for NGC 6992, we estimated the background contribution, obtaining a flux density measurement of 18.2$\pm$0.8~Jy and 9.8$\pm$1.8~Jy at 7.0 and 24.8~GHz, respectively.

\begin{figure*}
  \includegraphics[width=\columnwidth]{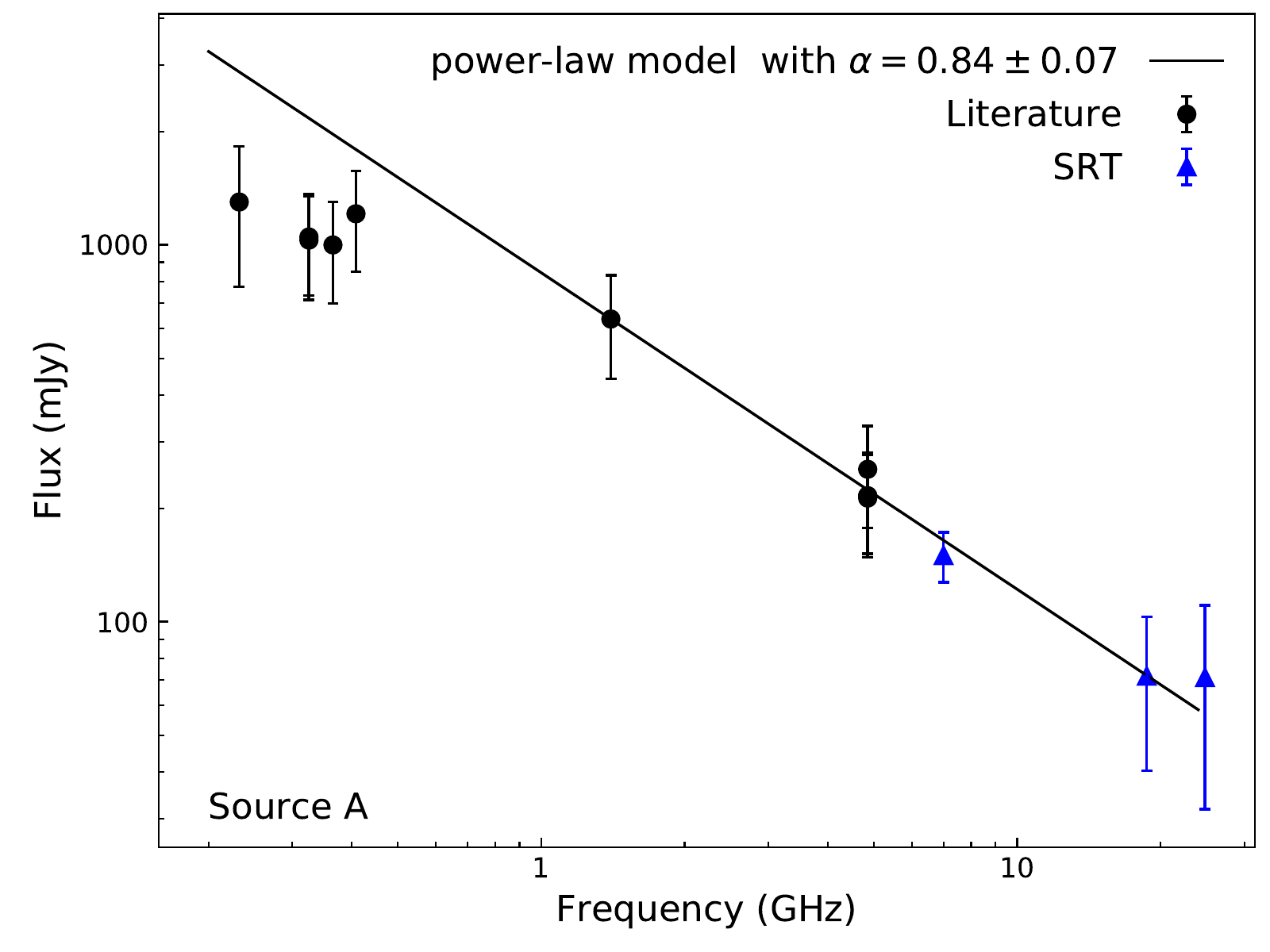}
  \hspace{-0.2cm}
  \includegraphics[width=\columnwidth]{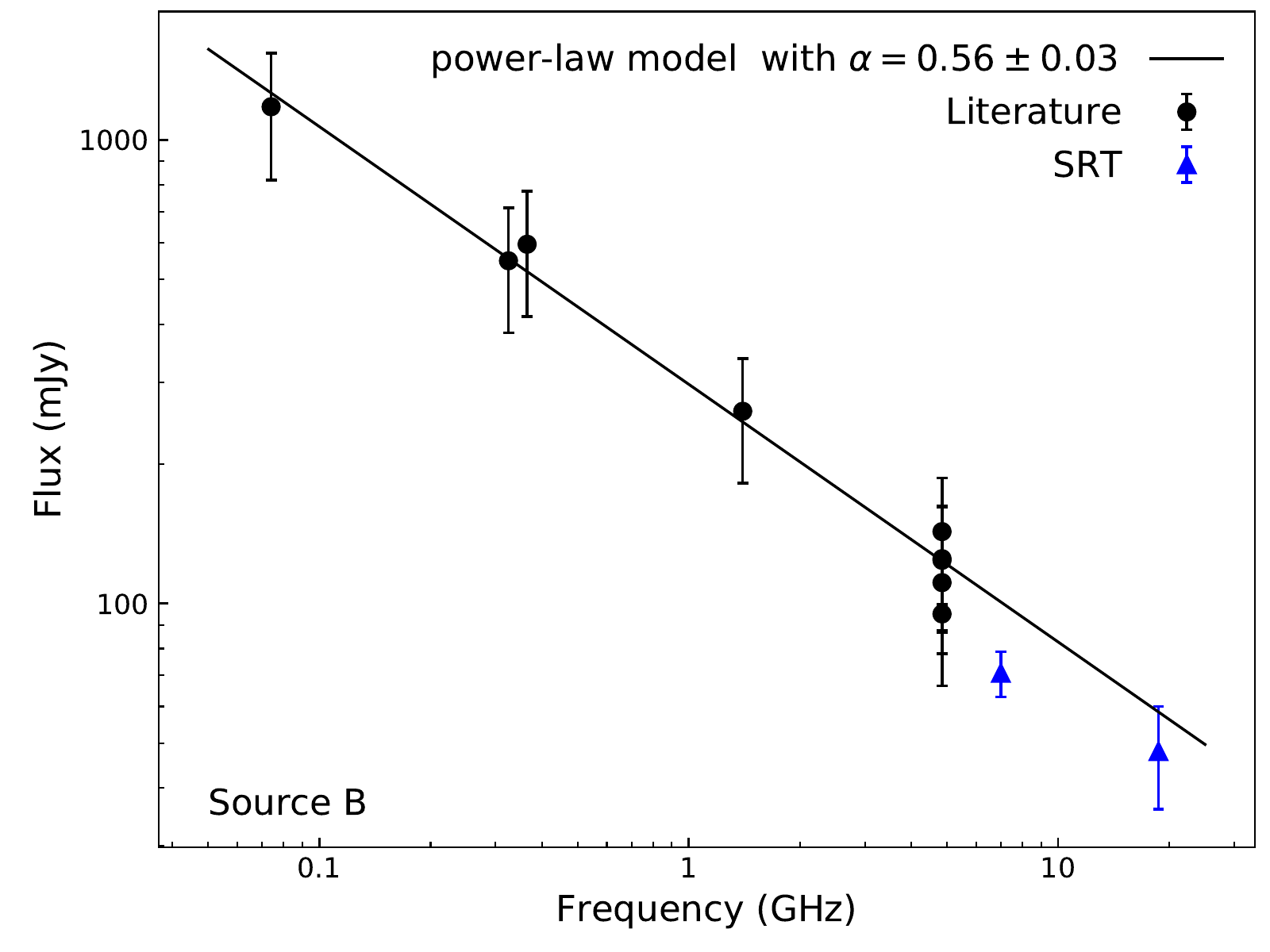}

\caption{SEDs of the extragalactic point sources `A' (NVSS J205800+314231, \textit{left}) and `B' (NVSS J205458+312614, \textit{right}) located near NGC 6992. The black dots indicate the literature measurements, and the blue triangles indicate the flux densities obtained in this work. The weighted least-squares fit applied to the source A is related to the 1.4-GHz and 4.85-GHz data, while that applied to the source B is associated with all the literature data.}
\label{SED_point_source_filament}
 \end{figure*}
 
 \begin{figure*}
  \includegraphics[width=\columnwidth]{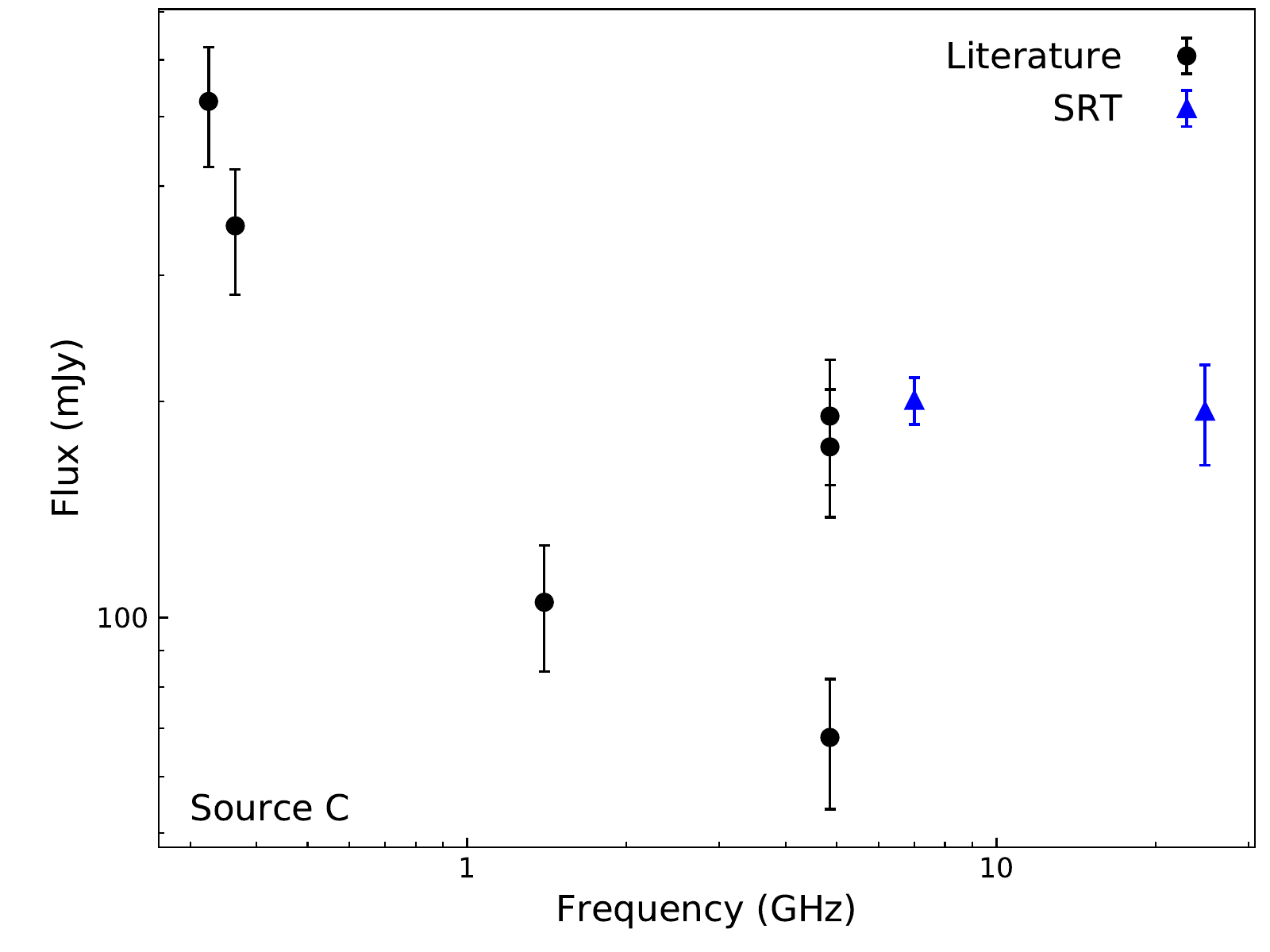}
  \hspace{-0.08cm}
  \includegraphics[width=\columnwidth]{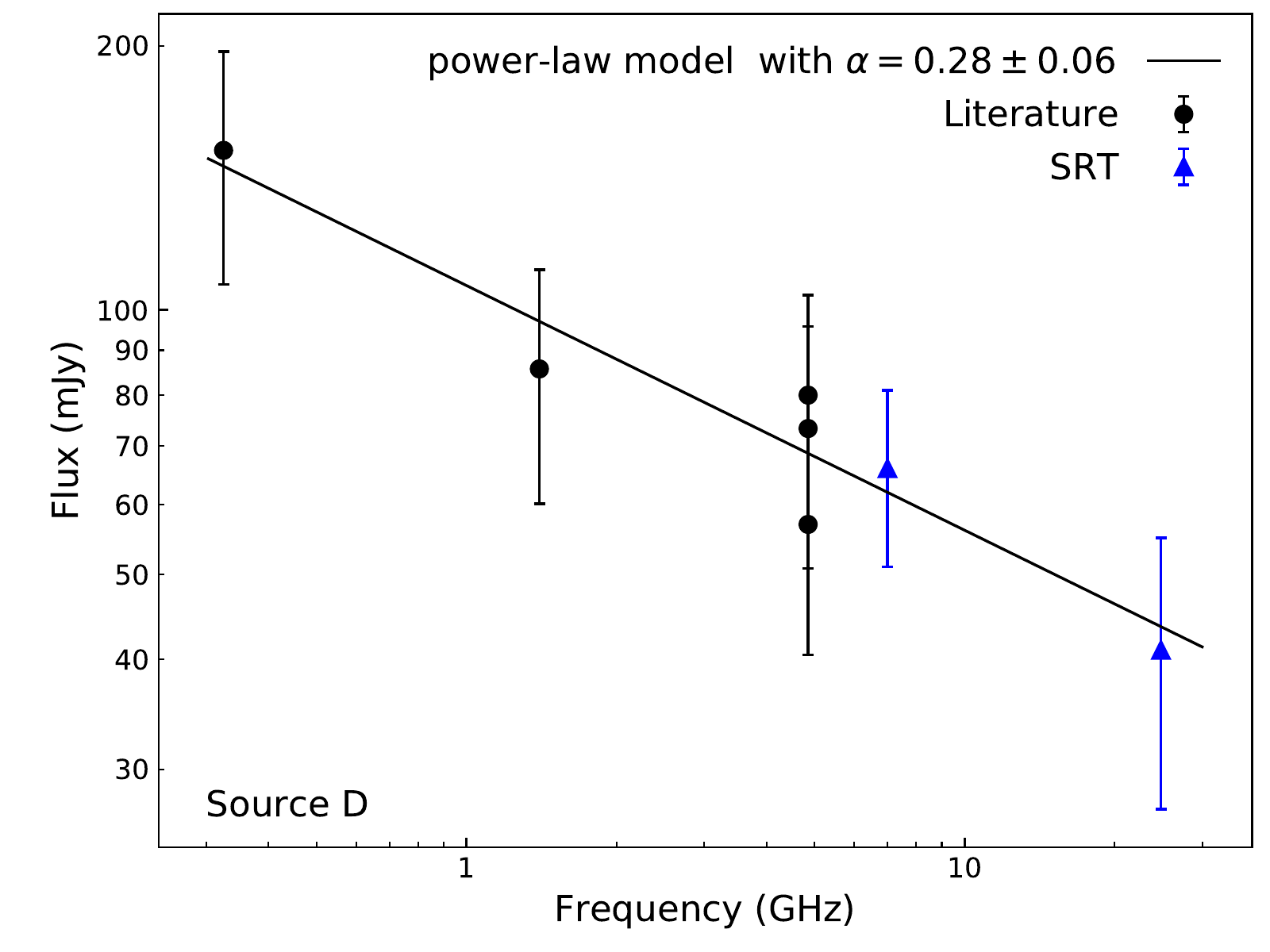}

\caption{SEDs of the extragalactic point sources `C' (NVSS J205122+291140, \textit{left}) and `D' (NVSS J205109+291629, \textit{right}) located inside the southern shell. The black dots represent the literature measurements, and the blue triangles represent the flux densities obtained in this work.
}
\label{SED_point_southern_shell}
 \end{figure*}

\subsection{Serendipitous point sources in the Cygnus Loop field}

Several compact radio sources were detected within the boundary of the Cygnus Loop or in its immediate vicinity (\citealt{Keen_1973}, \citealt{Green_catalogue_2019}). 
In the images at 7.0, 18.7 and 24.8~GHz, we detect four point sources located in close proximity to the remnant. These sources are, instead, confused within the Cygnus Loop emission on the 8.5-GHz map due to its lower resolution. 
When possible (depending on the variability of the source), we exploited the literature flux density measurements associated with these sources in order to cross-check our calibration and data analysis procedure.

Two point sources are well detected in the 7.0-GHz map of NGC 6992, and are indicated by the white-dashed circles in Fig.~\ref{CygB_maps}. These sources are reported in the NVSS catalogue \citep{Condon_1998} as NVSS J205800+314231 (`A' in Fig.~\ref{CygB_maps}) and NVSS J205458+312614 (`B' in Fig.~\ref{CygB_maps}), respectively. The source A is also detected at 18.7 and 24.8~GHz, while the weaker source B is well-detected only at 18.7~GHz. 
For both point sources, radio flux density measurements in the range $\sim$0.2$-$5~GHz are available on VizieR\footnote{\url{https://vizier.u-strasbg.fr/}}
along with their spectral energy distribution (SEDs). Although a classification of these sources is not available in the literature, the radio data and their location outside the Galactic plane suggest that they are likely radio galaxies.
We coupled the literature measurements with our flux densities of the source A at 7.0, 18.7 and 24.8~GHz and those at 7.0 and 18.7~GHz of the source B to assess the quality of our maps. 

The SEDs associated with the two sources are shown in Fig.~\ref{SED_point_source_filament}, where we also report the extrapolated fit of the spectra referred to the literature data. In the case of the source A, we excluded from the fit the data between 232 and 408~MHz, probably associated with a low-frequency spectral turn-over.
All the SRT data related to this source are perfectly consistent with the fit obtained from the literature data. In the case of the source B, our flux measurements at 7.0 and 18.7~GHz are consistent with the fit derived from the other data within 3$\sigma$ and 1$\sigma$, respectively.

In our 7.0-GHz image of the southern shell, we detected two point sources unrelated with the remnant, and indicated with white-dashed circles in Fig.~\ref{CygA_maps}. Differently from what was observed for NGC~6992, in the southern shell both point sources are superimposed on the remnant. The southern source (source `C' in Fig.~\ref{CygA_maps}) is listed in the NVSS catalogue as NVSS J205122+291140. Flux density measurements are available on VizieR
between 0.325 and 4.85~GHz.
This source is also included in the `Variable 1.4~GHz radio sources from NVSS and FIRST' catalogue \citep{Ofek_2011}. Indeed, as shown in Fig.~\ref{SED_point_southern_shell} (\textit{left}), the literature flux density measurements reveal a certain variability of this source, which is particularly highlighted by the spread of the NVSS flux densities at 4.85 GHz. Our measurements at 7.0 and 24.8 GHz also confirm this variability, which makes this source useless to cross-check our flux density measurements. 

The second point source (source `D' in Fig.~\ref{CygA_maps}) is named NVSS J205109+291629  \citep{Condon_1998}, and it is indicated as a blazar candidate in the `low-frequency radio catalogue of flat-spectrum sources' \citep{Massaro_2014}. The radio flux density measurements associated with this source are summarised in \cite{Vollmer_2010}, and the related SED is available on VizieR. 
The spectrum of this point source is shown in Fig.~\ref{SED_point_southern_shell} (\textit{right}). Both the SRT measurements are perfectly consistent with the fit derived from the archive data. The related spectral index is $\alpha = 0.28\pm0.06$, in agreement with the flat spectrum expected for this kind of objects.
It is worth to note that the flux densities of the source D at 7.0, 8.5 (extrapolated from the fit) and 24.8 GHz are within the uncertainties associated with the southern shell measurements at these frequencies (Sect.~\ref{Subsection: southern shell}). For this reason, we did not remove their contribution to the integrated flux density of the southern shell.

 \begin{table}
\noindent
	\centering
	\caption{Flux density measurements of the point sources observed close to NGC 6992 (sources A and B) and the southern shell (sources C and D)  with SRT between 7.0 GHz and 24.8 GHz. }
	\label{tab:Point sources flux densities}
	\begin{tabular}{|ccccc|} 
		\hline
		\hline

& & \multicolumn{3}{|c|}{Flux density (mJy)}\\
\cline{3-5}
&   &    \\

Source & ID & 7.0 GHz & 18.7 GHz & 24.8 GHz \\
\hline

NVSS J205800+314231 & A & 150.0 $\pm$ 23 & 72 $\pm$ 32 & 71 $\pm$ 39 \\

NVSS J205458+312614 &  B & 71 $\pm$ 8 & 48 $\pm$ 12& - \\
\hline
NVSS J205122+291140 & C & 201 $\pm$ 15 & - &  194 $\pm$ 31\\

NVSS J205109+291629 & D & 85 $\pm$ 20 & - & 41 $\pm$ 14 \\

 \hline
 \hline

	\end{tabular}
\end{table}

Our flux density measurements of the four point sources at 7.0, 18.7 and 24.8 GHz are given in Table~\ref{tab:Point sources flux densities}.
The derived spectra of the point sources A, B and D provide a useful verification of the reliability of our flux density measurements.

\section{Spectral analysis of the whole Cygnus Loop SNR}
\label{Sec: Cyg Loop spectral analysis}

 \begin{table*}
\noindent
	\caption{Integrated flux density measurements towards the whole Cygnus Loop SNR.}
	\label{tab:CygLoop flux densities literature}
    \hspace{-1.5cm}
	\begin{tabular}{|ccl|ccl|} 
		\hline
		\hline
Freq.	 & Flux density  & Reference & Freq.	 & Flux density& Reference   \\
 (GHz)	 & (Jy) &   & (GHz) & (Jy)& \\
  	 &  &   &   &  & 

  \\
 \hline
 \hline
 0.022 & 1378$\pm$400 &  \cite{Roger_1999} & 0.863  & 184$\pm$18  &  \cite{Uyaniker_2004} \\
 0.0345 & 1245$\pm$195 &  \cite{Sastry_1981} & 0.960 & 190$\pm$50 &  \cite{Kenderdine_1963} \\
 0.038&  956$\pm$150 & \cite{Kenderdine_1963}  & 1.420 & 143$\pm$14 &  \cite{Uyaniker_2004}\\
 0.041&  770$\pm$140 &  \cite{Kundu_1967_Cyg} & 2.675 & 115$\pm$12 &   \cite{Uyaniker_2004}\\
 0.158&  350$\pm$70 & \cite{Mathewson_1960}  & 2.695 & 125$\pm$16&  \cite{Green_1990} \\
 0.195 &   382$\pm$60 &  \cite{Kundu_1967_Cyg} & 2.7 & 88$\pm$6&  \cite{Kundu_1969}\\
 0.408&  230$\pm$50 & \cite{Kenderdine_1963}  & 4.940 & 73$\pm$7&  \cite{Kundu_1972}\\
 0.408&  260$\pm$50 &  \cite{Mathewson_1960} & 4.8 & 90$\pm$9 &  \cite{Sun_2006} \\
 0.408& 237$\pm$24 & \cite{Uyaniker_2004}& 8.5 & 54$\pm$4 & this work \\
 0.430  &  297$\pm$50 &  \cite{Kundu_1967_Cyg}  & 30 & 24.9$\pm$1.7 & \cite{Planck_2016} \\

\hline 
\hline
	\end{tabular}
\end{table*}

We analysed the radio spectrum of the entire Cygnus Loop SNR by coupling our measurement at 8.5 GHz and all the flux density measurements available in the literature. We also included the \textit{Planck} measurement at 30 GHz, the only one reported by \cite{Planck_2016} in Table 3 for this SNR.
All the values and related references are
reported in Table~\ref{tab:CygLoop flux densities literature}.

\begin{figure}
  \includegraphics[width=\columnwidth]{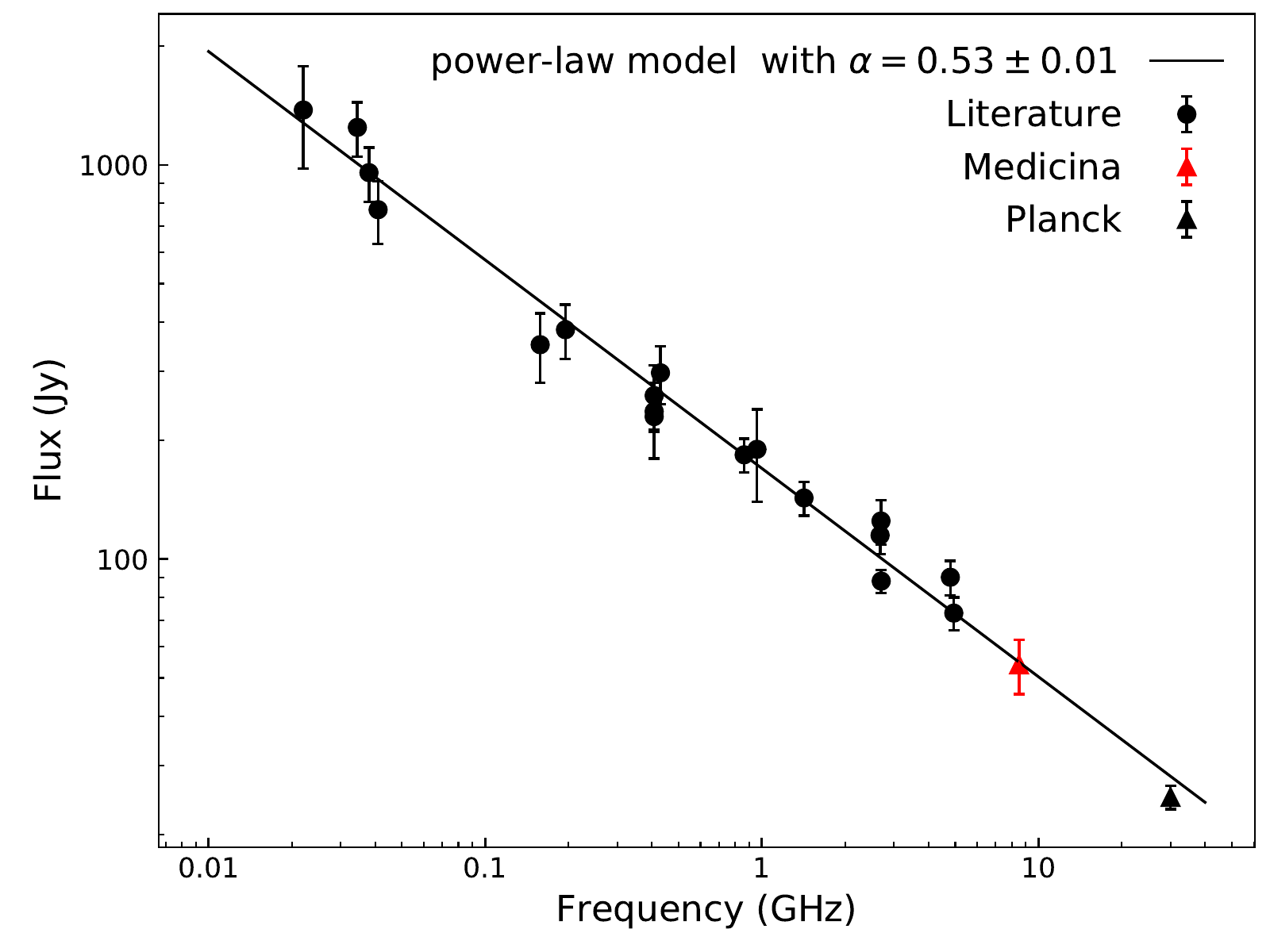}
\caption{Weighted least-squares fit applied to the Cygnus Loop
spectrum for the synchrotron power-law model. The red and black filled triangles correspond respectively to the Medicina point at 8.5 GHz
and the \textit{Planck} data at 30~GHz. All the flux density values and related references are given in Table~\ref{tab:CygLoop flux densities literature}.}\label{Cyg_SED}
 \end{figure}

It is worth noting that our measurement is the most sensitive obtained so far at these frequencies,
if we exclude the lower-resolution \textit{Planck} data. 
The overall radio spectrum is displayed in Fig.~\ref{Cyg_SED}. The Medicina value at 8.5~GHz is represented with a filled red triangle, and perfectly matches the tendency suggested by the other data without any apparent spectral variation.
We modelled the integrated spectrum of the Cygnus Loop using a simple synchrotron power-law function.
From the weighted least-squares fit ($\chi^2/dof = 1.15$), we obtained a spectral
index of $\alpha=0.53 \pm 0.01$ and a normalisation constant at 1~GHz of $164\pm5$~Jy. Our result is consistent within $1\sigma$ with the spectral index $\alpha=0.50\pm 0.06$ derived by \cite{Uyaniker_2004} by considering all the flux density measurements available in the literature between 0.022 and 4.94~GHz, and within $2\sigma$ with the spectral index
of $\alpha=0.42\pm0.06$ calculated by \cite{Uyaniker_2004} by considering the data from the Effelsberg 100-m telescope and the DRAO Synthesis telescope between 0.408 and 2.675~GHz, and that calculated by \cite{Sun_2006}, $\alpha=0.40\pm0.06$,
by adding the Urumqi telescope data at 4.8~GHz.
We also fit the data reported in Table \ref{tab:CygLoop flux densities literature} with a simple synchrotron model with an exponential cutoff:
\begin{equation*}
    S_{\nu}= K \left(\frac{\nu}{\nu_0}\right)^{-\alpha} e^{-\frac{\nu}{\nu_0}}
\end{equation*}
 where $K$ is the normalisation constant and $\nu_0$ is the cutoff frequency. The weighted least-squares fit ($\chi^2/dof = 1.09$) gives a spectral index $\alpha = 0.51 \pm 0.02$, $K = 167 \pm 5$~Jy (representing the flux density at 1~GHz) and a cutoff frequency $\nu_0 = 172 \pm 117$~GHz. Although the data are well-fitted by both the simple synchrotron power-law and the cutoff-frequency model, the high value of the obtained cutoff frequency suggests that the non-thermal synchrotron emission, without any steepening, is dominant in the considered frequency range.

Furthermore, our measurement rules out any spectral steepening up to high radio frequencies, and it confirms the tendency suggested by the \textit{Planck} data at 30 GHz.
Although the Cygnus Loop is considered approximately at the same evolutionary phase of other SNRs that showed a synchrotron spectral break, as in the case of S~147 \citep{Fuerst&Reich_1986} and W44 \citep{Loru_2018}, or characteristic spectral features, as the spectral bump observed in the IC443 between 20 and 70~GHz (\citealt{Onic_2017}, \citealt{Loru_2018}), its integrated spectrum appears dominated by the non-thermal synchrotron emission without any noticeable deviation. This result could indicate an earlier evolutionary phase of this SNR, for which a spectral steepening is expected at higher frequencies, or the global effect of different electron populations across the remnant undergoing peculiar shocks conditions.

\begin{table*}
  \caption{Summary of the parameters used to estimate the minimum energy ($U_{\rm min}$) and related magnetic field ($B(U_{\rm min})$) under the equipartition condition. We indicated with `sphere' and `shell' the source geometry assumed as a roughly uniform sphere or a spherical shell, respectively. 
  $V$, $L$ and $\alpha$ are the volume, the total luminosity and the spectral index, respectively.}
    \begin{tabular}{lcccccc}
\hline
\hline

  &\multicolumn{4}{c}{parameters} &\multicolumn{2}{c}{Results}   \\
 \cmidrule(lr){2-5} \cmidrule(lr){6-7}
Source & geometry & $V$ &  $L$ &  $\alpha$  & $U_{\rm min}$ & $B(U_{\rm min})$ \\
    &  & (cm$^3$) &  (erg s$^{-1}$)  & & (erg) & ($\mu$G)\\ \midrule
Cygnus Loop & sphere & $7.8 \times 10^{59}$ & $1.3 \times 10^{33}$ &  0.53 &  $2.8 \times 10^{49}$ & 19  \\
 & shell &  $2.9 \times 10^{59}$ & $1.3 \times 10^{33}$ & 0.53 &  $1.9 \times 10^{49}$ & 27   \\
NGC 6992 & - & $3.7 \times 10^{57}$ & $1.4 \times 10^{32}$ & 0.45 &  $7.3 \times 10^{47}$ &  46   \\
Southern shell & sphere & $1.3 \times 10^{59}$ & $3.1 \times 10^{32}$ & 0.49 &  $5.4 \times 10^{48}$ &   21 \\
 & shell-I & $3.3 \times 10^{58}$ & $3.1 \times 10^{32}$ & 0.49 & $ 3.0 \times 10^{48}$&  31 \\
  & shell-II & $1.9 \times 10^{58}$ & $3.1 \times 10^{32}$ & 0.49 & $2.4 \times 10^{48}$&  36 \\
\hline
\hline
\end{tabular}%

\label{tab: equpartition parameters}%
\end{table*}%
On the other hand, the presence of a synchrotron spectral steepening above $\sim$~20~GHz 
could be hidden by other processes, like thermal Bremsstrahlung or thermal dust-emission, which may become dominant depending on environmental conditions. On this base, we considered an upper cutoff frequency of 25~GHz and a lower cutoff frequency of 22~MHz, the latter corresponding to the lower-frequency flux density measurement available in the literature, to estimate the minimum total energy contents U$_{\rm min}$ in the Cygnus Loop and the related magnetic field strength B (U$_{\rm min}$). We here estimate these quantities under the assumption of the equipartition between particles and magnetic energy \citep{Essential_Radio_Astronomy}:
\begin{equation}
    U_{\rm min}=0.50 (\eta A L)^{4/7} V^{3/7} \mathrm{erg}
     \label{E eq}
\end{equation}
\begin{equation}
    B (U_{\rm min})=\left( \frac{6 \pi \eta A L }{V}\right)^{2/7} \mathrm{G}
    \label{B eq}
\end{equation}
where $V$ and $L$ are the volume and the total luminosity of the source expressed in cm$^3$ and erg s$^{-1}$, respectively;
$\eta$ is a factor that accounts for the contribution to the total energy of protons and heavier ions with respect to electrons; $A$ is a physical parameter that depends on the lower ($\nu_1$) and upper ($\nu_2$) cutoff frequencies and
the spectral index ($\alpha$):
\begin{equation*}
    A=\frac{C_1^{1/2}}{C_3} \frac{(2-2\alpha)}{(1-2\alpha)} \frac{\nu_2^{1/2-\alpha}-\nu_1^{1/2-\alpha}}{\nu_2^{1-\alpha}-\nu_1^{1-\alpha}} \quad \quad (\rm for \, \alpha \neq 0.5 \, \rm or \, 1)
\end{equation*}
where $\rm C_1$ and $\rm C_2$ are constants with value $\rm C_1 = 6.266  \times 10^{18}$ and $\rm C_3 = 2.368 \times 10^{-3}$ in cgs units \citep{Dubner_2015}. 
We assumed an isotropic emission in the frequency range from 22~MHz to 25~GHz and $\eta \sim 50$, as it was usually done in previous works on SNRs \citep{Castelletti_2007}. We assumed these values also in the case of NGC 6992 (Sect.~\ref{Subsection: NGC6992}) and the southern shell (Sect.~\ref{Subsection: southern shell}).
The other parameters are summarised in Table~\ref{tab: equpartition parameters} together with the results.
We performed the analysis for two geometrical models: i) by approximating the Cygnus Loop to an uniform sphere of radius $1.44$~deg; ii) by considering the CR electrons confined within the Cygnus Loop shell, with a thickness of approximately 5~per cent of the SNR radius.
The assumption of the energy equipartition between particles and magnetic field provides a first estimation of the minimum value of the energy and magnetic field strength in a SNR.
However, it is very difficult to decide whether synchrotron sources are in equipartition, especially in the case of extended and complex object like SNRs \citep{Dubner_2015}. Observational constraints from radio and $\gamma$-ray data are then needed to firmly constrain the particle energetics and the magnetic field strength, and to investigate on a possible SNR departure from the equipartition condition.
Indeed, the maximum energy achieved by an electron population with cutoff frequency $\nu_c$ is \citep{Reynolds_2008}:
\begin{equation}
    E = 14.7 \left( \frac{\nu_c/\mathrm{GHz}}{B/\mathrm{\mu G}}\right)^{\frac{1}{2}} \mathrm{GeV} \qquad
    \label{Obs eq.}
\end{equation}
Considering a tentative lower limit on the radio cutoff frequency
of 25~GHz and the maximum particle energy between 1 and 10~GeV, as derived from the $\gamma$-ray observations by \cite{Katagiri_2011}, we obtained a minimum value of the magnetic field ranging from 54~$\mu$G to 5.4~mG from Eq.~\eqref{Obs eq.}.

On the other hand, a stringent upper limit on the magnetic field can be obtained from the condition that the synchrotron cooling time has to be larger than the remnant's age, otherwise a cooling break should be visible in the radio spectrum. The cooling time is given by \citep{Ohira_2012}:
\begin{equation}
    t_{\rm sync} = \frac{9 m_e^2}{4 r_0 c B^2 E} 
    = 1.1 \times 10^{9} \, \left(\frac{B}{\mu G} \right)^{-3/2} \, \left(\frac{\nu_c}{\rm GHz}\right)^{-1/2} \; {\rm yr} \,,
\end{equation}
where $m_e$ and $r_0$ are the electron mass and the classical electron radius, respectively. In the second equality, we substituted $\nu_c$ from Eq.~\eqref{Obs eq.}, Hence, imposing $t_{\rm sync} > t_{\rm age}$ with the condition $\nu_c > 25$\,GHz, we get $B < 470\,\mu$G.

A more constraining investigation on the particle energetics and magnetic field strength can be achieved through a  model of the radio and $\gamma$-ray emission from the Cygnus Loop, which accounts for the electron radiative losses and the evolution of the magnetic field, coupled to the SNR dynamics.
A detailed description of this method is provided in Sect.~\ref{sec:model}.

\section{Spectral analysis of NGC 6992 and the southern shell}
\label{Sec: Spectral analysis of NGC 6992 and the southern shell}

In this Section, we use the Medicina and SRT data and the flux density measurements at 30 and 40 GHz, derived from the public \textit{Planck} maps, in order to investigate the integrated spectrum of the two regions of the Cygnus Loop: NGC 6992 and the southern shell.
Due to the lack of separated $\gamma$-ray flux measurements for these two regions, we cannot use the Eq.~\eqref{Obs eq.} or the model described in Sect. \ref{sec:model} to constrain the magnetic field strength and the particle spectrum. We will estimate, instead, the magnetic field using only radio data under the assumption of energy equipartition between non-thermal particles and magnetic field.
 \begin{table}
\noindent
	\centering
	\caption{Flux density measurements and rms related to the maps of NGC 6962 and the southern shell  carried out with the Medicina radio telescope and SRT. }
	\label{tab:CygLoop flux densities}
	\begin{tabular}{|cccc|} 
		\hline
		\hline
Source	 & Freq.  &  Flux density & $\sigma$  \\
 name  & (GHz) & (Jy)& (mJy/beam)\\
  \\
 \hline
 \hline
 
\textbf{NGC 6962} & 7.0 & 8.7 $\pm$ 0.6 & 15.8\\
                  & 8.5 & 7.5 $\pm$ 0.9  & 43.7\\
                  & 18.7 & 5.6 $\pm$ 0.7  & 3.4\\
                  & 24.8 & 4.5 $\pm$ 0.9 & 8.9 \\

\hline 
\textbf{Southern shell}  & 7.0 & 18.2 $\pm$ 0.8 & 6.5 \\
                         & 8.5 & 16.6 $\pm$ 1.5   & 44.7\\
                         & 24.8 & 9.8 $\pm$ 1.8 & 6.6 \\

\hline 
\hline
	\end{tabular}
\end{table}
 \begin{table}
\noindent
	\centering
	\caption{Flux density measurements of the regions NGC 6962 and the southern shell related to the \textit{Planck} maps of the  Cygnus Loop at 30 and 44 GHz. }
	\label{tab:Planck flux densities}
	\begin{tabular}{|ccc|} 
		\hline
		\hline
Source	 & Freq.  &  Flux density   \\
 name  & (GHz) & (Jy)\\
  
 \hline
 \hline
 
\textbf{NGC 6962} & 30 & 4.32 $\pm$ 0.4 \\
& 44 & 4.57 $\pm$ 0.9 \\

\hline 
\textbf{Southern shell}  & 30 & 9.49 $\pm$ 0.5 \\
& 44 & 8.25 $\pm$  1.1 \\


\hline 
\hline
	\end{tabular}
\end{table}
\begin{figure*}
  \includegraphics[width=8.2cm]{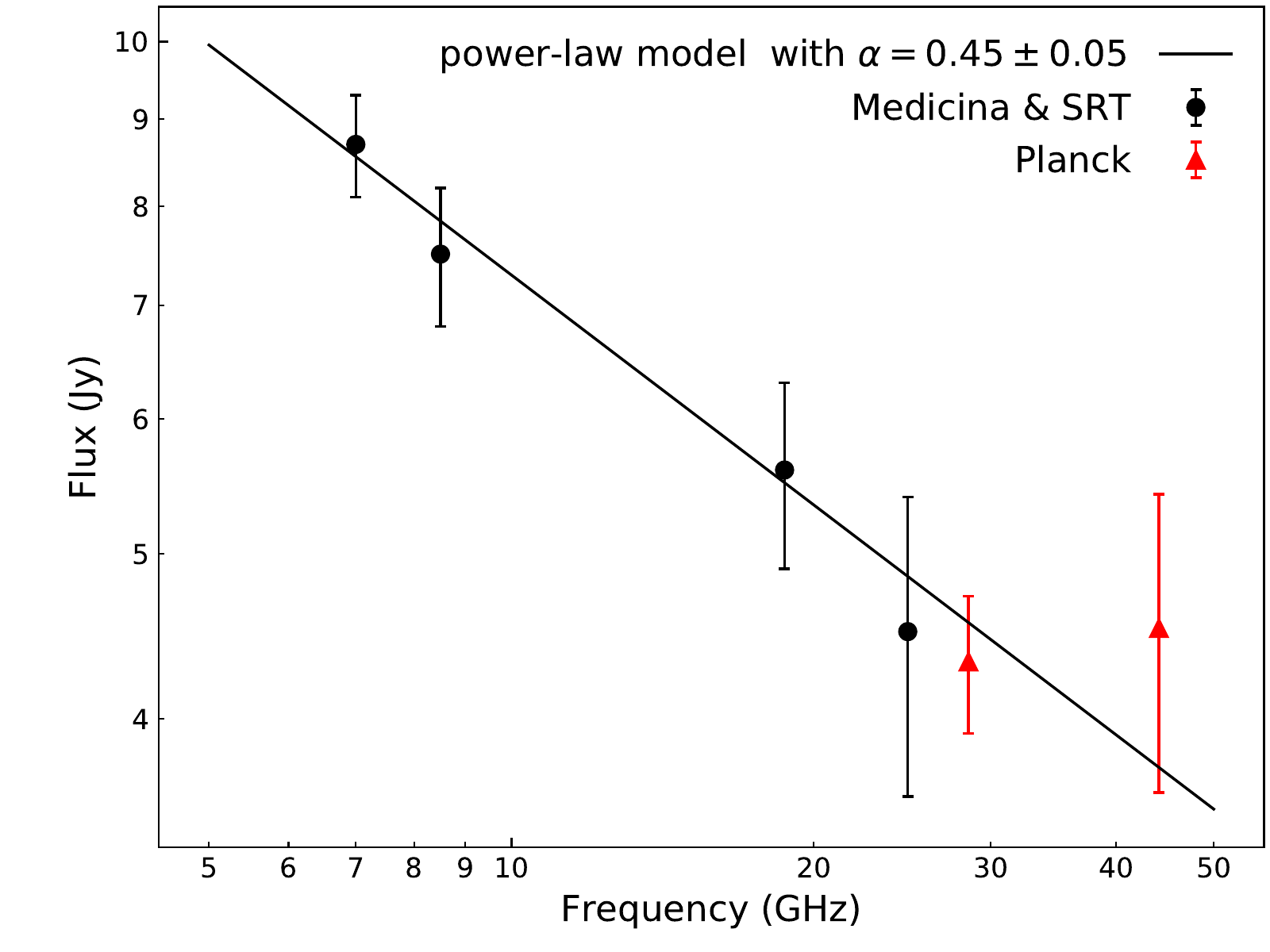}
 \hspace{-0.1cm}
 \includegraphics[width=8.2cm]{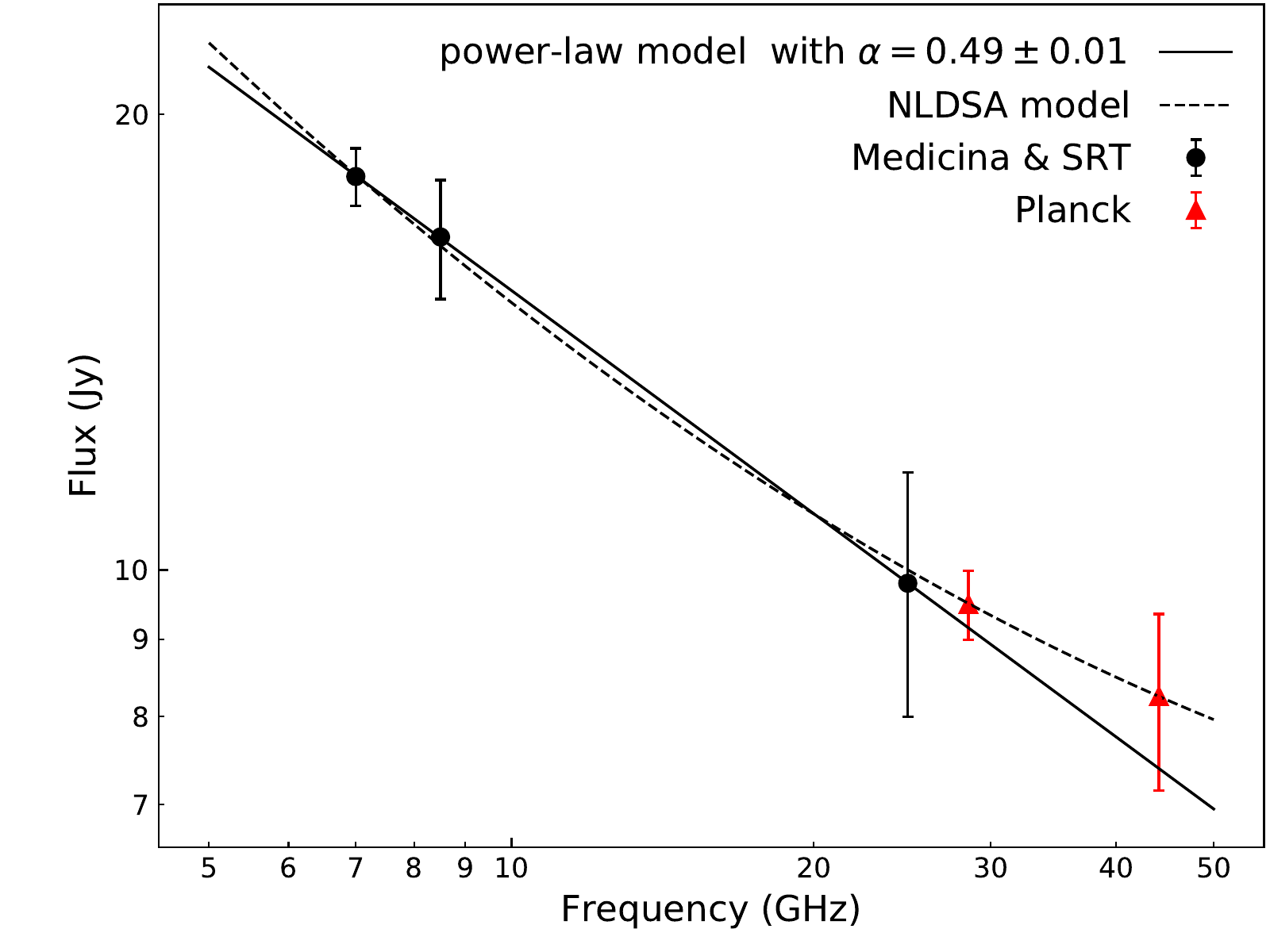} 
\caption{\textit{Left}: weighted least-squares fit applied to our measurements at 7.0, 8.5, 18.7 and 24.8 GHz, indicated with black circles, of NGC 6992 by using the synchrotron power-law model. 
The red triangles correspond to the \textit{Planck} data at 30 and 44 GHz. \textit{Right}: spectral energy distribution related to the southern shell. The black solid line shows the weighted least-squares fit applied to our flux density measurements at 7.0, 8.5 and 24.8 GHz for a simple  synchrotron power-law model. The non-linear diffusive shock acceleration model is represented by a black dotted line and it is obtained tacking also the \textit{Planck} measurements at 30 GHz and 44 GHz (red triangles) into account}\label{CygB_SED}
 \end{figure*}
 
\begin{figure*}
  \includegraphics[width=9.0cm]{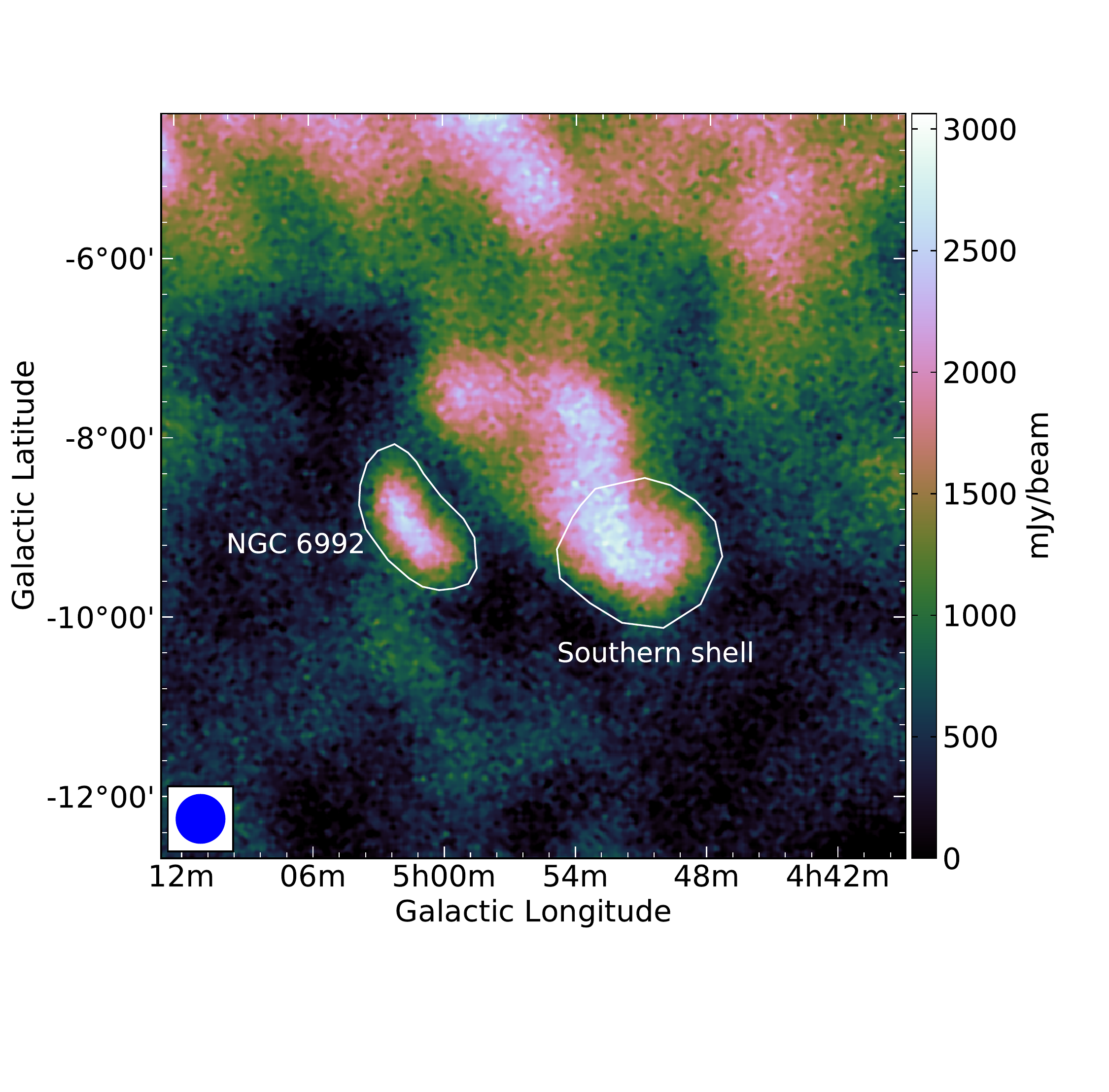}
  \hspace{-0.5cm}
  \includegraphics[width=9.0cm]{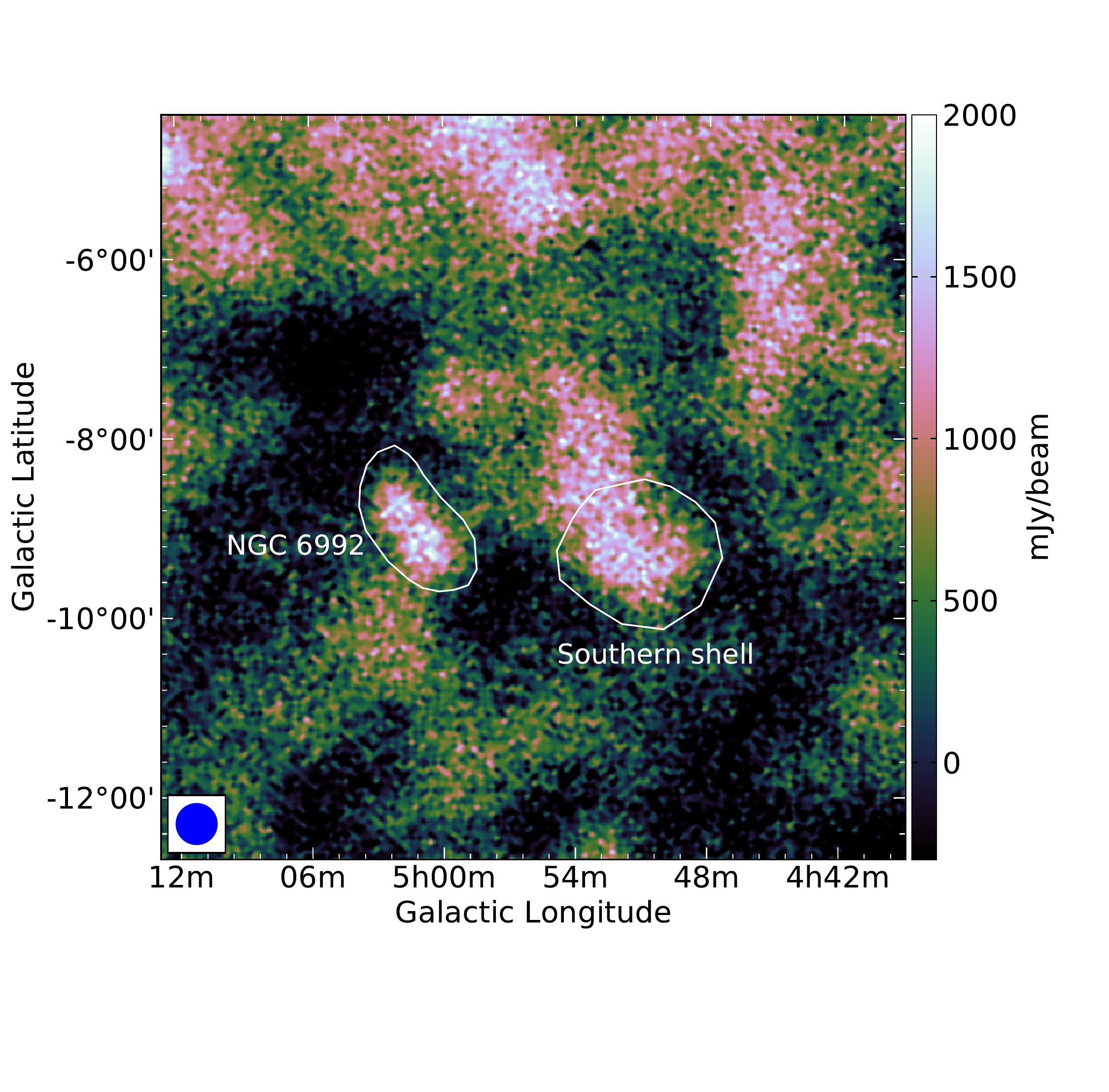}

\caption{The Cygnus Loop images from \textit{Planck} at 30 GHz (\textit{left}) and 44 GHz (\textit{right}). The beam size of 32.3~arcmin at 30~GHz and 27.1~arcmin at 44~GHz is indicated by the blue-filled circle on the bottom left corner of the maps. 
The white contours indicate the extraction regions that we used to calculate the flux density. }
\label{Cygnus_Loop_Planck}
 \end{figure*}

\subsection{NGC 6992}
\label{Subsection: NGC6992}
The morphology of NGC~6992 is governed by the interaction of the SNR blast wave with smaller discrete clouds \citep{Fesen_2018}.
The correlation between X-ray and optical emission reflects the non-homogeneous interaction of the blast wave with the clouds: in the low-density inter-cloud medium, the shock wave propagates unimpeded and emits X-ray radiation, while it decelerates where it encounters dense clumps of gas with a resultant optical emission \citep{Levenson_1998}.
Local changes in shock velocity and in the preshock density were also pointed out by optical studies of the H$\alpha$ filaments observed in NGC 6992 \citep{Blair_2005}.
These result in isolated regions
where the non-radiative shock is becoming radiative.
NGC 6992 corresponds to the brightest sector of the $\gamma$-ray Cygnus Loop emission observed with \textit{Fermi-LAT} between  1 and 100~GeV  \citep{Katagiri_2011}, suggesting the interaction of high-energy particles with the circumstellar medium (CSM) regions.

We used the flux densities calculated from our maps at 7.0, 8.5 and 18.7 GHz to investigate the integrated spectrum associated with NGC 6992. These values are listed in Table \ref{tab:CygLoop flux densities}. We performed a weighted fit of our data by using a simple power-law model, as shown in Fig.~\ref{CygB_SED}. We calculated for this region $\alpha=0.45\pm0.05$, which is significantly lower ($>1\sigma$) than the value that we obtained for the whole Cygnus Loop SNR ($\alpha$ = 0.53 $\pm$ 0.01). 
This result confirms and extends up to high radio frequencies the slightly flatter spectra ($\alpha \simeq 0.4$) revealed for this region by the spatially-resolved spectral index studies performed by \citet{Uyaniker_2002} in the frequency range from 0.408 to 2.675~GHz.

There are several explanations for flat radio spectra of SNRs in the current literature \cite[see][for an overview]{Onic_2013}. Most of the models involve a significant contribution of the second-order Fermi mechanism, but some of them also discuss high compression ratio, contribution of secondary electrons produced in hadronic collisions, as well as the possibility of thermal contamination. Given the relatively low density in the region of NGC 6992 ($\sim 5$\,cm$^{-3}$), contributions from secondary electrons and thermal emission are disfavoured. 

Concerning the second-order Fermi acceleration, we have considered the model developed by \cite{Ostrowski_1999}, which includes this mechanism in the shock acceleration and predicts hard spectra when the ratio $\rm v_A/u_{\rm sh}\gtrsim 0.1$ (where $\rm v_A$ is the Alfv\'en velocity and $u_{\rm sh}$ is the shock velocity) and/or the particle diffusion is much flatter, in energy, than the Bohm-like diffusion. However, in the case of NGC 6992, a spectrum with $\alpha=0.45$ requires either an upstream magnetic field $B_1 \approx 80\,\mu$G, a value which we will exclude in our next analysis (see the last part of this Section and Sect.~\ref{sec:model}), or a diffusion coefficient $D\propto E^{0.5}$. The latter condition is also at odds with the idea that diffusion results form the self-generated turbulence. In such a case, particle spectra harder than $E^{-2}$ would produce self-generated diffusion harder than $E^1$, contradicting the premise. Hence a different turbulence origin should be invoked.

We are left with the possibility of large compression. Such a condition is naturally realised in radiative shocks because the downstream plasma loses energy through radiation and becomes more compressible. \cite{Blair_2005} found that some portions of the shock in NGC 6992 are close to become radiative. The same conclusion was reached by \cite{Szentgyorgyi2000} for the region NGC 6995. Then it is possible that the compression ratio ($r$) is just slightly larger than 4. 
Remembering that the radio spectral index is connected to the shock compression ratio as $\alpha = 3/(2(r-1))$, we need $r=4.3$ to produce $\alpha= 0.45$. Such an interpretation is very attractive, but comes with some caveats. First of all, radiative shocks are not thought to be efficient accelerators even though it is not very clear when the transition from efficient to inefficient acceleration occurs. Indeed, the Cygnus Loop may represent a rare example where such a transition may be studied. Secondly, when a shock becomes radiative, its velocity structure is more complex than the simple step function used to describe high-velocity shocks. Therefore, the final particle spectrum should be computed using the correct velocity profile but this is beyond the aim of the present work.

We used the \textit{Planck} public maps of the Cygnus Loop available on the NASA/IPAC Infrared Science Archive\footnote{\url{https://irsa.ipac.caltech.edu/Missions/planck.html}} to compute the flux density of NGC~6992 at 30 and 44~GHz. We considered the same extraction region as shown in Fig.~\ref{Cygnus_Loop_Planck}. The background-corrected results are listed in Table~\ref{tab:Planck flux densities}.
We calculated the errors associated with the \textit{Planck} measurements adding in quadrature the statistical uncertainty and the flux calibration uncertainty. For the latter, we assumed the values of the mission calibration uncertainty reported in the Table~1 of \cite{Planck_2016}.
The \textit{Planck} flux density at 30 GHz is consistent with the tendency suggested by our data. This represents a further confirmation that no spectral steepening takes place in this region up to high radio frequencies. The value at 44 GHz indicates, instead, a significant flux density increase, suggesting that the dust emission contribution becomes important. On the other hand, we cannot exclude a possible contribution of the dust emission also at lower frequencies, in particular on the flux density at 30 GHz. This could compete with the synchrotron emission, hiding a possible spectral cutoff. 
However, we note that our high-frequency maps, especially that at 24.8 GHz, are affected by artefacts and high noise, which together with the weakness of the source emission make a precise flux density estimation difficult. More $K$-band SRT observations are needed to firmly establish the NGC~6992 spectral behaviour at high radio frequencies.

As in the case of the whole Cygnus Loop,
we established a lower limit for a possible spectral break in the NGC 6992 region at $\sim$25 GHz and used it for estimating the minimum energy and related magnetic field by assuming that this region satisfies the equipartition condition. The used parameters and results are summarised in Table~\ref{tab: equpartition parameters}. 
The magnetic field value (46~$\mu$G) is in perfect agreement with the estimation performed by \cite{Sun_2006} for the same region considered in the energy equipartition condition.
On the other hand, the magnetic field obtained for NGC 6992 is higher than that obtained for the entire Cygnus Loop SNR both modelling it as an uniform sphere (19~$\mu$G) or a shell (27~$\mu$G), suggesting a possible magnetic field amplification process in this region like the turbulent dynamo.

\subsection{Southern shell}
\label{Subsection: southern shell}

The southern shell is the brightest region of the Cygnus Loop in the radio domain. It is a typical shell centred at about ($\alpha$, $\delta$) = (20$^h$49.56$^m$, 29$^{\circ}33'$) with a size of $\sim1.4^{\circ}\times1.8^{\circ}$ \citep{Uyaniker_2002}. 
Radio polarimetric studies revealed a tangential configuration of the magnetic field along the whole southern shell, as typically expected for a middle-aged SNR in its adiabatic phase \citep{Sun_2006}.
Spatially-resolved spectral index studies performed on the whole Cygnus Loop by \citet{Leahy_1998} between 0.408 and 1.420~GHz revealed
region-dependent spectral indices, ranging from flat spectra corresponding to the north and northeast rims, to steep spectra in the southern shell and in the faint region on the west of the central bright filament of the northern shell. By including also the radio data at 2.695~GHz, the authors also pointed out region-dependent spectral shapes with a negative curvature (spectral steepening) associated with NGC 6992 and a positive curvature (concave-up) in the central faint regions of the northern shell and in the southern shell.
More recent observations by \citet{Uyaniker_2004} between 0.408 and 2.675~GHz showed slight variations of the southern shell spectral index compared to the integrated value of the whole Cygnus Loop SNR ($\alpha \sim0.5$). In particular a steeper spectrum was observed in the central fainter region ($\alpha \sim0.6$), while lower spectral indices are found in the brightest filaments ($\alpha \sim0.4$). The authors attributed these variations to a weak compression between the shock wave and the ISM in the southern shell, where the shock acceleration appears as the dominant mechanism. 

Optical observations revealed a fully radiative region on the east side of the shell. However, the filamentary structures are less extended than those observed in the northern shell, suggesting smaller clouds or a more recent interaction of the SNR shock with them \citep{Levenson_1998}. The southern shell is very faint in the X-ray images, with an extremely smooth shell detected in the soft X-ray band \citep{Levenson_1999}. 
The radio continuum and the $\gamma$-ray emission do not appear correlated in this region
\citep{Katagiri_2011}.
Only the bright southern $\gamma$-ray spot (centred at $\sim$ $\alpha$, $\delta$=$20^h53^m$,$+29^{\circ}20'$) is coincident with the brightest radio region of the shell. 

We investigated the high-frequency spectrum of the southern shell by using the flux densities at 7.0, 8.5 and 24.8~GHz. The resulting spectrum is shown in Fig.~\ref{CygB_SED} ($right$), where we also included the \textit{Planck} measurements at 30 GHz and 44 GHz (see Table~\ref{tab:Planck flux densities}). 
Between 7.0 and 24.8~GHz,
the spectrum results perfectly fitted by a simple power-law function with $\alpha=0.49\pm0.01$, ruling out any spectral steepening.
Our fit underestimates the \textit{Planck} data. If we include them in our analysis, the spectrum results well fitted by a synchrotron power-law function with $\alpha=0.44\pm0.02$. The uncertainties related to our flux density at 24.8 GHz make it compatible with this fit. Nevertheless, both \textit{Planck} values seem to suggest a concave-up deviation from the simple synchrotron-emission model. We tried to investigate the emission processes and the theoretical models that could be at the basis of this spectral tendency. \\`Concave-up' radio spectra, flattening to high frequencies, were observed both in evolved (composite, mixed-morphology) and young shell-like SNRs. In the first case, the curvature of the spectrum is related to other emission mechanisms that become significant at high frequencies, especially for SNR evolving in complex environmental conditions \citep{Urosevic_2014}. The most accredited models include bremsstrahlung emission in presence of dense ISM environment, thermal dust emission linked to cold dust in molecular clouds and spinning dust emission from very small grains usually found in nearby Galactic molecular and dust clouds interacting with the SNR shocks.  
The density in the southern shell is even lower than the one in NGC 6992 \citep{Fesen_2018}, hence a possible significant contribution from these processes is not expected.
In particular, the thermal dust contribution, expected at these frequencies, is not strongly evident at 30 and 44 GHz. It could become significant at higher frequencies. 
In some young SNRs, in the free-expansion and in the early adiabatic-expansion phases of their evolution \citep{Dubner_2015}, a curved concave-up shape is observed as a possible result of the dynamical reaction of accelerated particles that modify the shock structure. Under the modified shock configuration, the low-energy electrons are confined in the subshock region where they undergo a compression factor $r<4$. This results in a steepening of the radio spectra at low frequencies. On the contrary, because the high-energy electrons can sample a more extended space region ($r>4$ far from the shock) the related energy spectrum results harder (flat radio spectrum). The overall effect is a concave-up radio spectrum due to a significant CR production and confinement (\citealt{Amato-Blasi2005}, \citealt{Urosevic_2014}). The magnetic field configuration of the southern shell is consistent with that of a shell-like SNR in the adiabatic phase, and it results different from that of the northern shell, where the magnetic field shell structure appears more irregularly \citep{Sun_2006}. 
The nature of the southern shell is still debated. On the basis of neutral Hydrogen line observations, \citet{Leahy_2005} interpreted this region as a result of a southern cavity arising from the Cygnus Loop expansion on a wall of neutral gas with consequent acceleration of smaller interstellar clouds adjacent to 
the northern rim of this region toward the center of the Cygnus Loop.
On the other hand, in the scenario of two SNRs composing the Cygnus Loop, the radio and magnetic field characteristics  are attributed to an early evolutionary stage of the southern SNR, which would have formed after the northern one \citep{Uyaniker_2004}. The absence of a spectral steepening, due to a still very efficient acceleration mechanism up to high radio frequencies, could be a confirmation of this hypothesis. In this context, the indication of a concave-up spectrum could be associated with an efficient CR production at the shock.

In order to investigate this possibility, we fitted our data and the \textit{Planck} measurements at 30 GHz and 44 GHz by a model that includes  non-linear diffusive shock acceleration (NLDSA) effects. We represented this model with a varying power-law of the form $S_{\nu}\propto \nu^{-\alpha-c \log\nu}$, where $c$ is a curvature parameter.
The resulting fit for the varying spectral-index model is represented by the dotted line in Fig.~\ref{CygB_SED} (\textit{right}).
We obtained a spectral index  $\alpha = 0.72 \pm 0.02$, a spectral curvature parameter  $c = 0.052\pm 0.003$ and a normalisation constant of 61$\pm$2 Jy at 1~GHz. The spectral index is significantly higher compared to that obtained for the simple synchrotron power-law model,
indicating a flattening with increasing frequencies. 
Our estimation of the curvature parameter is perfectly consistent with that obtained by \cite{Onic_Urosevic_2015} ($c = 0.056 \pm 0.008$) for the case of the young SNR Cas~A, which they ascribed to an efficient non-linear diffusive shock accelerator. Furthermore, the  spectral index obtained for the southern shell is consistent within $2\sigma$ with that obtained for Cas~A ($\alpha=0.790 \pm 0.016$).
We point out that our flux density measurements, including the \textit{Planck} ones, perfectly agree with both the simple synchrotron emission and the non-linear synchrotron model due to the large uncertainties, and do not allow us to constrain the spectral tendency in the frequency range between 20 and 50~GHz.
Therefore, we cannot completely exclude a possible dust emission contribution, although not dominant.
More flux-density measurements with higher sensitivity between $\sim8$ GHz and $\sim50$ GHz could be crucial to constrain the particle-acceleration mechanisms in the southern shell.

Assuming the southern shell in the energy-equipartition condition, we calculated the minimum energy and associated magnetic field strength by using Eqs. \ref{E eq} and \ref{B eq}. In Table~\ref{tab: equpartition parameters}, we list the used parameters and related results. We assumed  two shell models with a thickness of: i) $\sim$5~per cent of the Cygnus Loop radius; ii) $\sim$5~per cent of the southern shell radius. 
We observed that the magnetic field strength of the southern shell, considered as an uniform sphere ($B(U_{\rm min}) = 21~G$), is very close to that obtained for the whole Cygnus Loop, suggesting very similar conditions ($B(U_{\rm min}) = 19~G$). 
On the other hand, the estimation performed by considering the southern shell as an independent SNR (shell-II) provides a higher magnetic field ($B(U_{\rm min}) = 36~G$) compared to that obtained for the entire Cygnus Loop. This could be consistent with an amplification of the magnetic field as observed in SNRs with a concave-up radio spectrum and resulting from an efficient shock acceleration that modifies the shock structure  \citep{Reynolds_2011}.
However, only a detailed modelling of the non-thermal emission from the radio to the $\gamma$-ray band will firmly constrain the magnetic field strength in this region.

\section{Modelling the non-thermal emission} 
\label{sec:model}
In this Section, we model the non-thermal emission from the whole Cygnus Loop, combining radio and $\gamma$-ray observations, to investigate the properties of particle acceleration at the forward shock and their time-dependent escape from the source.

One of the main issues in the shock acceleration theory is identifying the maximum energy that particles can reach and understanding how it evolves in time. This is also intimately connected with the magnetic field amplification, a  process thought to occur in the shock region by means of the same accelerated particles and a necessary requirement to reach energies much larger than a few tens of GeV \cite[for recent reviews see][]{Blasi2013Rev,Amato2014Rev}. To this respect, middle-aged SNRs can be very helpful given that several among these remnants show a $\gamma$-ray spectrum with a cutoff, or a spectral softening above $\sim 1-10$\,GeV, and the Cygnus Loop makes no exception \citep{Katagiri_2011}. Such a feature could be related to the maximum energy of the particles accelerated at the present time \citep{Celli+2019,Brose+2020}. In fact, the break, or softening, could be produced by particles accelerated in the past that are no more confined by the acceleration mechanism and start diffusing away from the source in such a way that a fraction of escaped particles are still located inside the source.
If this were indeed the case, a spectral break is expected in the radio band at a frequency given by Eq.~\eqref{Obs eq.}, namely $\nu_{\rm br} \simeq 4.6 \,(E_{\max,e}/{10 \rm ~GeV)}^2 \, (B/{\rm 10 ~\mu G)}$\,GHz. In turn, if the $\gamma$-ray emission is due to hadronic processes, the maximum energy inferred from the high-energy cutoff rather refers to protons, such that
the magnetic field remains generally poorly constrained and $\nu_{\rm br}$ cannot be determined. Hence, it is fundamental to describe both contributions from accelerated hadrons and leptons, and related radiative emissions, to obtain a clear understanding about the relative importance of the many ongoing physical processes.
For this reason, we will use a self-consistent model, where the magnetic field at the shock is estimated from the maximum energy of particles through the amplification process. Here, the temporal evolution of the SNR is taken into account to calculate adiabatic losses for both particles and magnetic field, in order to correctly determine the whole synchrotron spectrum.

With such a time-dependent model, we can constrain several quantities like particle maximum energy, acceleration efficiency, ratio between accelerated electrons and protons, and magnetic field strength at different stages of the remnant evolution.
In order to keep the calculations as simple as possible, we will use a spherically symmetric model, which in turn will not allow to account for the different radio slopes observed in different parts of the Cygnus Loop.

\subsection{Particle acceleration model}
\label{sec:mod_theory}
The model that we adopt here is developed by  \cite{Celli+2019} and \cite{Morlino-Celli2020}, hence the reader is referred to those papers for further details. 
For simplicity, we assume that the remnant is expanding into a uniform CSM, with numerical density $n_0$. The SNR properties (namely SN explosion kinetic energy $E_{\rm SN}$, circumstellar density $n_0$, SNR age $t_{\rm age}$ and distance from us $d$) are fixed to the best values obtained by \cite{Fesen_2018}, and summarised in Table~\ref{tab:parameters}. We then follow \cite{Truelove_McKee99} to describe the SNR evolution assuming an ejecta power-law index equal to 0 \cite[see Table 5 in][]{Truelove_McKee99}. The ejecta mass is then fixed to $5\,M_{\odot}$ to match the measured shock proper motion of 0.1~arcsec~yr$^{-1}$, which at $d=735$~pc, corresponds to a shock velocity $u_{\rm sh}= 367~\rm km\,s^{-1}$ \citep{Fesen_2018}. At the same time, the predicted shock radius, $R_{\rm sh}= 20$\,pc, well matches with the E-W semi-axis of 18.5~pc.

We account for particle acceleration at the forward shock assuming that a fixed fraction of the shock kinetic energy is transferred to non-thermal particles. The instantaneous proton spectrum accelerated at the shock is
\begin{equation} 
\label{eq:f_0}
 f_{p,0}(p,t) = \frac{3 \, \xi_{\rm CR} u^2_{\rm sh}(t) \rho_0}{4 \pi \, c (m_p c)^4  \Lambda(p_{\max}(t))} 
 		\left( \frac{p}{m_p c} \right)^{-s}  e^{ -p/p_{\max}(t)} \,,
\end{equation}
where $\rho_0=n_0 m_p$ ($m_p$ being the proton's mass), $\Lambda(p_{\max}(t))$ is a normalisation constant such that the CR pressure at the shock is $P_{\rm CR} = \xi_{\rm CR} \rho_0 u_{\rm sh}^2$. The factor $\xi_{\rm CR}$ represents the instantaneous acceleration efficiency, and it is kept constant during the whole evolution of the SNR until now. We also remember that the linear diffusive shock acceleration process (DSA) predicts the particle spectral slope to be $s=4$ for strong shocks.
$p_{\max}(t)$ represents the instantaneous maximum momentum (and $E_{\max}$ is the corresponding maximum energy) achieved at the time $t$. Simple considerations on particle confinement suggest that it increases during the ejecta dominated (ED) phase and decreases during the Sedov-Taylor (ST) phase, following a simple power-law: 
\begin{equation} 
\label{eq:pmax0}
 p_{\max}(t) =
  \begin{cases} 
   p_\textrm{M} \left( t/t_{\rm Sed} \right)     & \text{if } t \leqslant  t_{\rm Sed} 	\\
   p_\textrm{M} \left( t/t_{\rm Sed} \right)^{-\delta}     & \text{if } t > t_{\rm Sed}
  \end{cases}
\end{equation}
where $t_{\rm Sed}\simeq 1500$\,yr is the beginning of the ST age. Both the absolute maximum momentum $p_{\rm M}$ and the slope $\delta$ are free parameters, that will be constrained from data. We also stress that the final result is not very sensitive to the behaviour of $p_{\max}$ for $t < t_{\rm Sed}$, as a consequence of the fact that $t_{\rm Sed} \ll t_{\rm age}$. Hence the behaviour of the maximum energy during the ED phase cannot be constrained.

Differently from protons, electrons are affected by radiative losses, so that we need to estimate the magnetic field strength in order to determine their distribution at the shock. To be conservative, we only evaluate the magnetic field at the shock required to reach the maximum energy. The common assumption is that this field is self generated by the same particles through resonant \citep{Skilling1975} or non-resonant instabilities \citep{Bell_2004} \cite[see also][]{Amato-Blasi2009}. Here, we do not specify the precise mechanism, but we make the assumption that the magnetic turbulence has a flat power distribution to produce a Bohm-like diffusion coefficient in the upstream of the shock, i.e. $D_1 = r_L(p) c /(3\mathcal{F}(k_{\rm res}))$ where $r_L= p c/(e B_0)$ is the particle's Larmor radius and $\mathcal{F}(k)$ is the logarithmic power spectrum of magnetic turbulence related to particle momentum through the resonant condition $k_{\rm res} = 1/r_L(p)$. In the following, $B_0$ refers to the unperturbed circumstellar magnetic field, while $\delta B$ represents the turbulent component, so that the total magnetic field upstream reads as $B_1=\sqrt{B^2_0+\delta B_1^2}$. In addition, subscript 1 (2) refers to quantities evaluated upstream (downstream) of the shock. When the magnetic field is amplified beyond the linear regime ($\delta B \gg B_0$), the diffusion becomes Bohm-like in the amplified magnetic field \citep{Blasi2013Rev}. In other words, the function $\mathcal{F}$ should have the following limits: $\mathcal{F} \sim (\delta B/B_0)^2$ for $\delta B \ll B_0$  and $\mathcal{F} \sim (\delta B/B_0)$ for $\delta B \gg B_0$. 
A minimal formula that reproduces these limits is  $\mathcal{F} = \left[(B_0/ \delta B) + (B_0/ \delta B)^2\right]^{-1}$, which, once inverted, gives
\begin{equation}
\label{eq:deltaB_F}
  \delta B_1(t) = \frac{B_0}{2} \left( \mathcal{F}(t) + \sqrt{4 \mathcal{F}(t) + \mathcal{F}^2(t)} \right) \,.
\end{equation}  
The relation between the maximum energy and the turbulent magnetic field upstream, $\delta B_1$, is obtained by imposing that the maximum energy is determined by the age of the SNR, i.e. $t_{\rm acc}=t_{\rm age}$. Using the acceleration time from quasi-linear theory, namely $t_{\rm acc} \simeq 8 D_1(p_{\max})/u_{\rm sh}(t)^2$ \citep{Morlino_2017}, we have $\mathcal{F}(t) = 8 r_L(p_{\max}(t)) c/(3 u_{\rm sh}(t)^2 t)$.
This equation allows us to compute the magnetic turbulence self-generated by accelerated particles at the shock, once a receipt is given for the maximum momentum and the shock speed temporal evolution.
The magnetic field upstream is then compressed at the shock and expands adiabatically during the SNR evolution.

The electron spectrum is similar to that of protons, but the maximum energy and the cutoff shape are different. In the loss-dominated case, i.e. for $t_{\rm sync}<t_{\rm age}$, a super-exponential cutoff is expected \citep{Zirakashvili-Aharonian:2007,Blasi:2010}. In particular, when energy losses are proportional to $E^2$, like in synchrotron and inverse Compton (IC) processes, the spectral cutoff is proportional to $\exp{[-(p/p_{\max,e})^2]}$. A good approximation to the spectrum is provided by \cite{Zirakashvili-Aharonian:2007}:
\begin{equation} 
\label{eq:f_e0_2}
 f_{e,0}(p) = K_{\rm ep}\,f_{p,0}(p)  {\left[1+0.523 \left(p/p_{\max,e}\right)^{\frac{9}{4}}\right]}^2 \, e^{- \left(\frac{p}{p_{\max,e}} \right)^2} \,,  \\
\end{equation}
where the constant $K_{\rm ep}$ accounts for the different injection efficiency of electrons with respect to protons. The electron maximum momentum, $p_{\max,e}$ is determined (as a function of time) either by losses or by the acceleration time, i.e. $t_{\rm acc}= \min[t_{\rm loss}, t_{\rm age}]$ where losses include both synchrotron emission and IC scattering. The latter is evaluated adopting the average Galactic background photons due to the cosmic microwave background (CMB), plus infrared, optical and UV radiation produced by dust emission and star-light.

Once proton and electron spectra are known at the shock, their evolution inside the SNR is calculated assuming that all particles with energies $E < E_{\max,p}(t)$ remain confined into their plasma elements, suffering adiabatic and radiative losses (for electrons). On the contrary, when $E > E_{\max,p}(t)$, particles (both protons and electrons) start escaping from the shock with a rate determined by the properties of the local CSM. Inferring such properties is quite difficult, hence we assume that particles diffuse with a local diffusion coefficient given by $D_{\rm csm}=\chi D_{\rm Gal}$, where $D_{\rm Gal}$ is the average Galactic diffusion coefficient as estimated from direct CR measurements \cite[see e.g.][]{evoli+2019} and $\chi$ is a free parameter that can be estimated from  observations.

\subsection{Results}
\label{sec:mod_results}
%
\begin{table*}
  \caption{Value of parameters used to model the Cygnus Loop spectrum. The left block refers to the SN explosion kinetic energy ($E_{\rm SN}$), the SNR age ($t_{\rm age}$) and distance ($d$), and the circumstellar density ($n_{0}$) as inferred by \citet{Fesen_2018}. The ejecta mass ($M_{\rm ej}$) is, instead, chosen to match the present values of the shock radius ($R_{\rm sh}$) and the shock velocity ($u_{\rm sh}$). The right block refers to the parameters used in the acceleration model described in Sect. \ref{sec:model}, where: $\xi_{\rm CR}$ is the acceleration efficiency; $s$ is the particle spectral slope; $E_{\rm M}$ is the absolute maximum proton energy; $\delta$ is the maximum momentum spectral slope; $K_{\rm ep}$ is a constant that accounts  for  the  different  injection efficiency  of electrons  with  respect  to  protons; $B_0$ is the unperturbed circumstellar magnetic field; $\chi$ is a free parameter related to the local diffusion coefficient.}
\makebox[\textwidth][c]{
    \begin{tabular}{cccccccccccccc}
\hline
\hline
      & \multicolumn{6}{c}{Cygnus Loop properties} & \multicolumn{7}{c}{Acceleration model parameters} \\
      \cmidrule(lr){1-7}
      & \multicolumn{3}{c}{Assumed} & \multicolumn{5}{c}{Derived} & \multicolumn{3}{c}{} \\
\cmidrule(lr){1-5} \cmidrule(lr){6-7}\cmidrule(lr){8-14}
$E_{\rm SN}$   &   $M_{\rm ej}$ &  $t_{\rm age}$  &  $d$        &  $n_{0}$       & $R_{\rm sh}$  & $u_{\rm sh}$ &  
$\xi_{\rm CR}$ &   $s$          &  $E_{\rm M}$    &  $\delta$   &  $K_{\rm ep}$  & $B_0$  &    $\chi$ \\ \midrule
$7\times 10^{50}$\,erg & 5 $M_{\odot}$ & $2.1\times 10^4$\,yr & 735\,pc & 0.4 cm$^{-3}$ & 20 pc & 380 km s$^{-1}$ & 0.07 & 4.0 & 200\,TeV & 3 & 0.15 & $3\,\mu$G & 1  \\
\hline
\hline
\end{tabular}%
}
\label{tab:parameters}%
\end{table*}%

\begin{figure}
    \centering
    \includegraphics[width=\columnwidth]{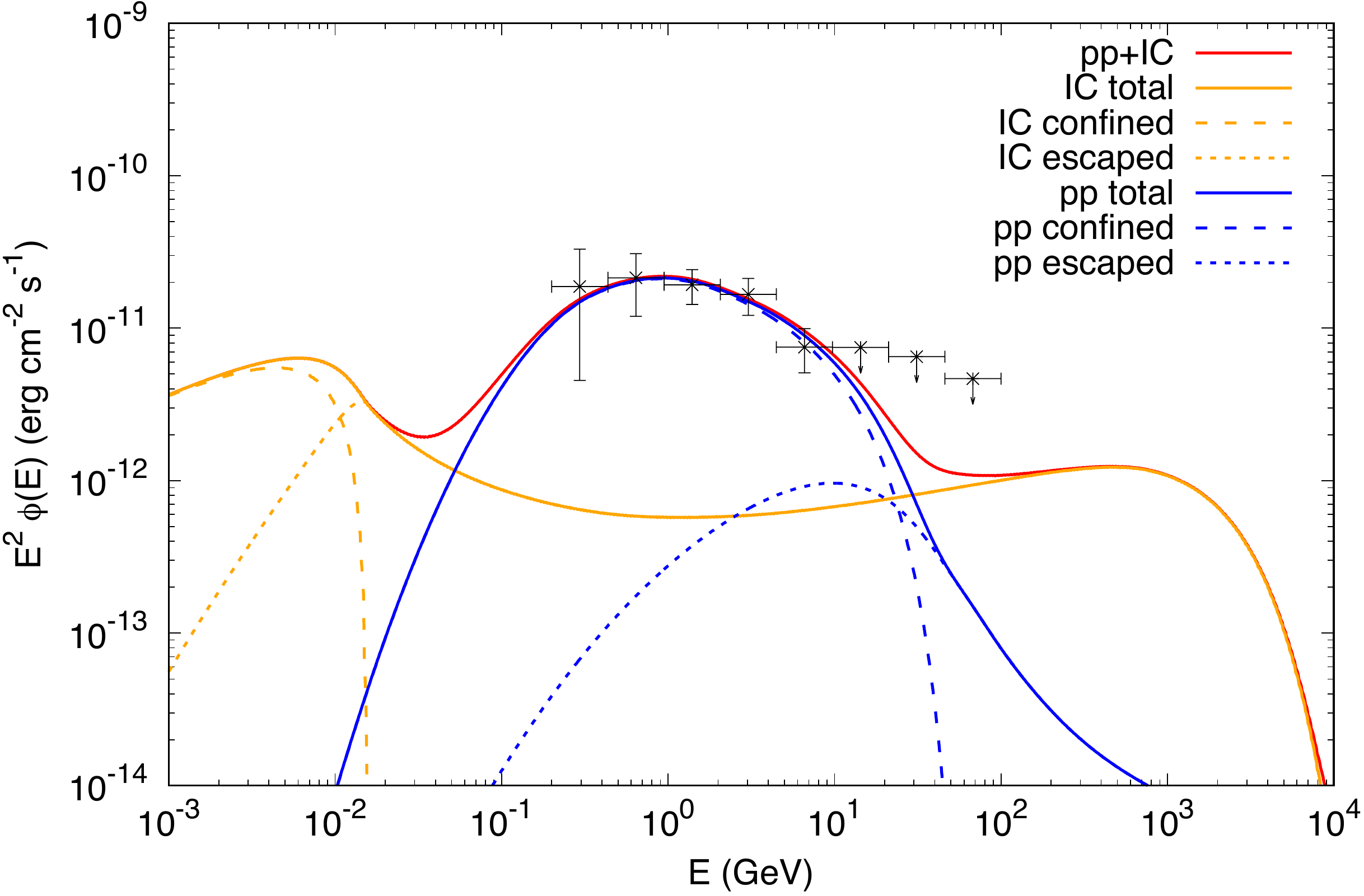}
    \caption{$\gamma$-ray emission estimated from the model compared with Fermi-LAT observations \citep{Katagiri_2011}. Hadronic (pp) and leptonic (IC) components are shown respectively with blue and yellow lines, while their sum is given in red. For each process, dashed and dotted lines show the contribution from confined and escaped particles, respectively.}
    \label{fig:gamma_flux}
\end{figure}

\begin{figure}
    \centering
    \includegraphics[width=\columnwidth]{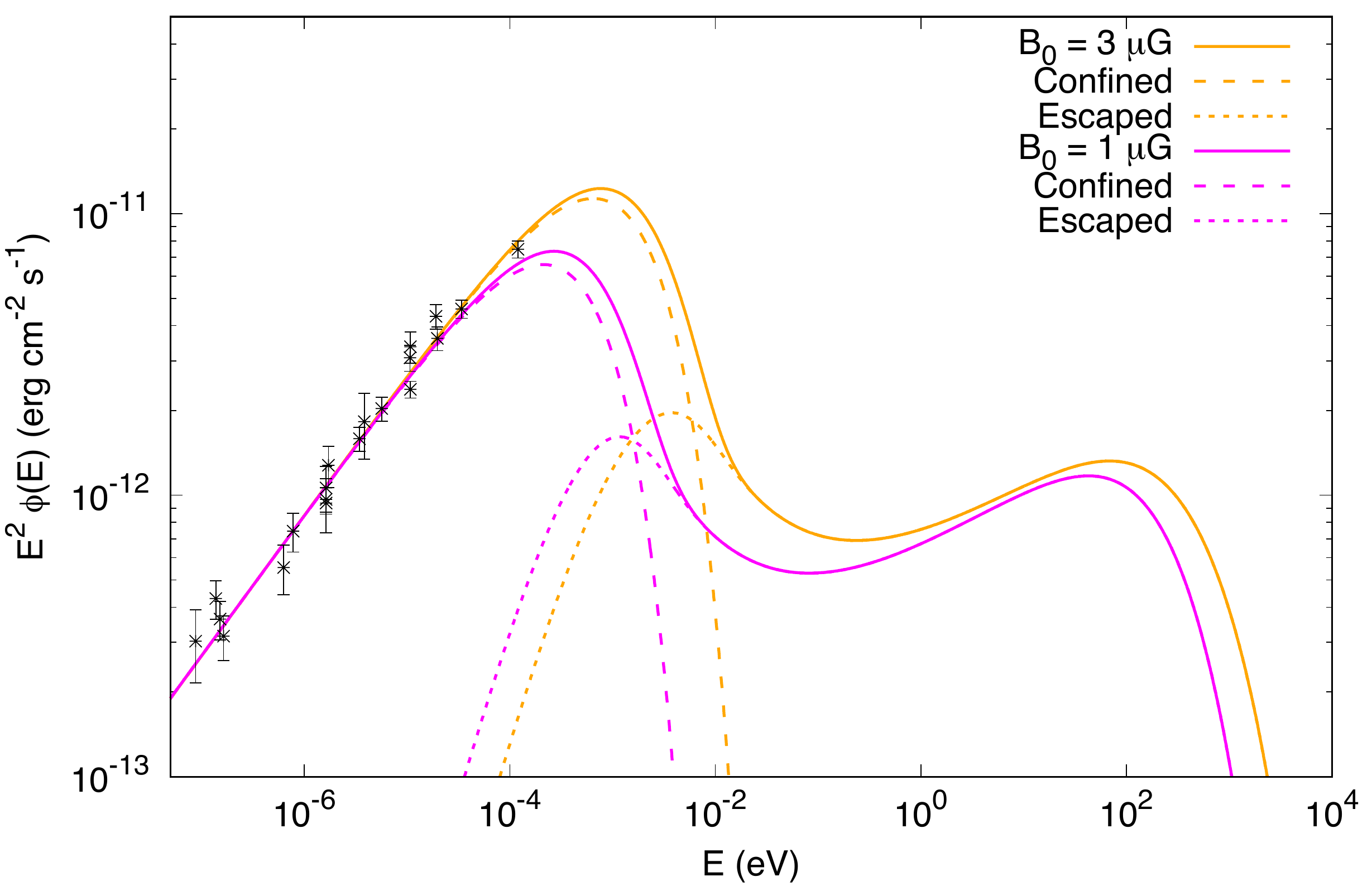}
    \caption{Synchrotron emission as derived from the model compared with radio data from the whole SNR. Yellow lines show our best model, which assumes $B_0= 3\,\mu$G, while the magenta lines show the case with $B_0 = 1\,\mu$G. Dashed and dotted lines show the contribution from confined and escaped electrons, respectively.}
    \label{fig:radio_flux}
\end{figure}

The model outlined in the previous Section has seven free parameters, listed in the right side of Table~\ref{tab:parameters}, which can all be constrained with a reasonable accuracy using radio and $\gamma$-ray observations. 
Figures~\ref{fig:gamma_flux} and \ref{fig:radio_flux} show the $\gamma$-ray and the radio emission, respectively, resulting from our best model, as compared with available data. The modelling result accounts for the radiation emerging from several mechanisms, including proton collisions on target gas density (pp interaction), electron IC scattering on the background photons and synchrotron emission from electrons in the SNR magnetic field. The $\gamma$-ray emission from pp interaction is calculated using the parametrization provided by \cite{Kafexhiu-Taylor2014}. 
Each radiation process contains two different contributions, one produced by confined particles and one from escaping particles.

\begin{figure}
    \centering
    \includegraphics[width=\columnwidth]{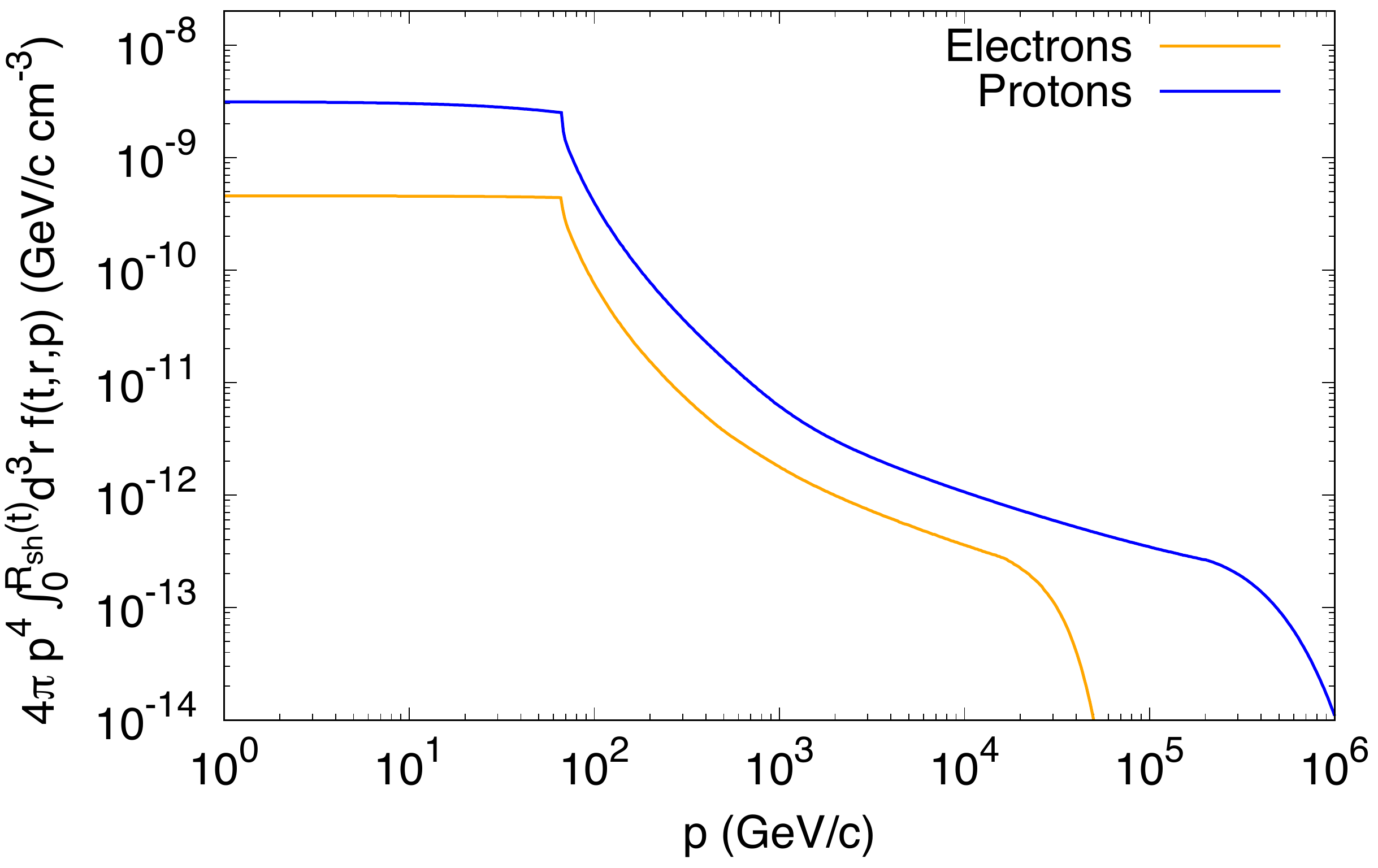}
    \caption{Accelerated proton and electron spectra located inside the SNR at the present age, integrated inside the whole SNR and multiplied by $4\pi p^4$. The break located at $65$\,GeV corresponds to the maximum energy reached at the present time hence, particles above this energy are escaping.}
    \label{fig:ep_spectra}
\end{figure}

To better understand the shape of the non-thermal radiation, it is useful to have a closer look at the particles spectra. Figure~\ref{fig:ep_spectra} shows both electron and proton spectra spatially integrated in the remnant interior at the present time. The maximum energy reached at $t_{\rm age}$ is 65\,GeV for both electrons and protons: below this energy, particles are all confined and their spectrum is proportional to $p^{-4}$, as set through Eq.~\eqref{eq:f_0}, while above the spectra steepen as a consequence of the escaping process. We highlight that the flattening observed towards the largest energies is due to the contribution of the shock precursor, where particles do not suffer adiabatic losses. The cutoffs observed at the highest energies are due to the absolute maximum energy reached at the beginning of the ST phase, which for protons is  $E_{M,p} \simeq 200$\,TeV while, for electrons is a factor 10 lower due to severe synchrotron losses, resulting in $E_{M,e} \simeq 20$\, TeV.
We note that the maximum energy we derived here is much larger than the interval $1-10$ GeV obtained by \cite{Katagiri_2011}.
{}
{}
{}
\begin{figure}
    \centering
    \includegraphics[width=\columnwidth]{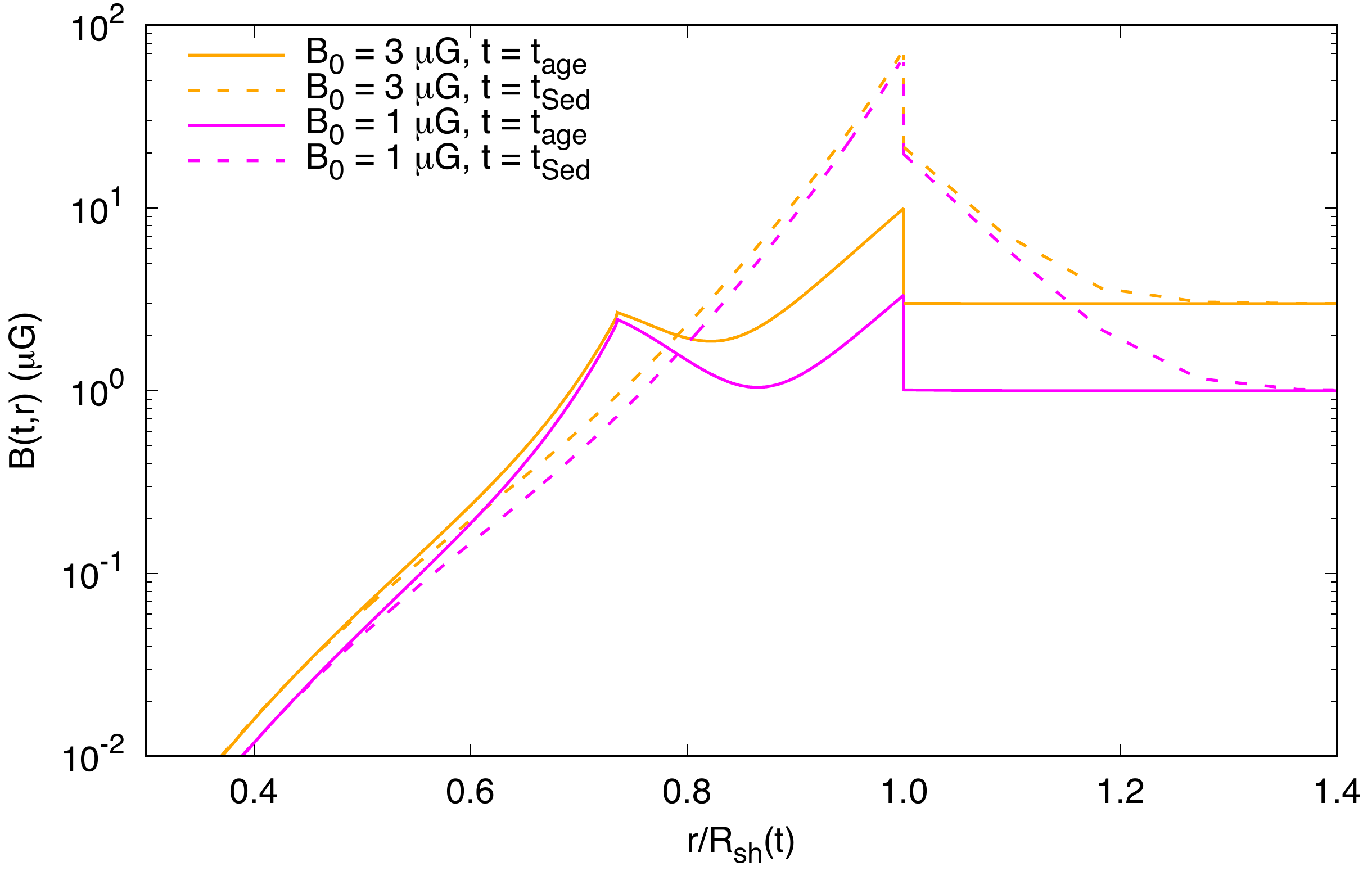}
    \caption{Total magnetic field strength inside and outside the SNR as predicted by the model presented in Sect. \ref{sec:mod_theory}, assuming $B_0=3\,\mu$G (upper-yellow lines) and $1\,\mu$G (bottom-magenta lines). Solid and dashed curves refer to $t=t_{\rm age}$ and $t=t_{\rm Sed}$, respectively. 
    The radial coordinate is divided by $R_{\rm sh}(t)$, which increases with time.}
    \label{fig:mag_field}
\end{figure}{}

In the following, we outline how data constrain each single parameter of the model.
First of all, the radio spectral index fixes the particle spectral slope through the relation $s = 2\alpha + 3$, hence $\alpha\simeq 0.5 \Rightarrow s\simeq 4$, in good agreement with linear DSA prediction.
As the observed $\gamma$-ray spectrum is expected to arise entirely though pp collisions, its normalisation allows us to fix the acceleration efficiency $\xi_{\rm CR}$, once the target density is fixed, while the spectral shape beyond few GeV simultaneously constraints $p_{M,p}$, $\delta$ and $\chi$. Unfortunately, the lacks data beyond a few 10~GeV results in some degeneracy between those three parameters.
Hence, we decided to make the conservative assumption $D_{\rm csm}=D_{\rm Gal}$, which allows us to constrain $p_{M,p}\simeq 200$\,TeV and $\delta\simeq 3$ in order to reproduce the $\gamma$-ray data as shown in Figure~\ref{fig:gamma_flux}.
It is worth stressing that the IC emission spectrum from escaping electrons is quite flat and dominates the emission above $\sim 30$ GeV. Even if those electrons have been accelerated in the past, a fraction of them is still located inside the SNR and its amount depends on the diffusion coefficient like $N_{\rm e} \propto D_{\rm csm}^{-3/2}$. As a consequence, our choice of $\chi=1$ should be considered as a lower limit, in that $\chi\ll 1$ would result in overshooting the Fermi-LAT upper limits above 10\,GeV.

An interesting test for our model would be to look for the TeV emission produced through IC by escaping electrons. The differential flux at 1 TeV is $\sim 10^{-12}$~erg s$^{-1}$ cm$^{-2}$, well within the sensitivity range of current imaging atmospheric Cherenkov telescopes (IACTs). Unfortunately, the Cygnus Loop is very extended, what makes the observation very challenging for the small field of view (FoV) of IACTs. An attempt was made by the MAGIC collaboration \citep{MAGIC-CL-2012}, resulting only in upper limits (compatible with our predictions).
The large IC flux produced by the escaping electrons in the $\gamma$-ray band results from the large electron density as inferred from the radio data. In fact, radio data allow to estimate the remaining parameters, $K_{\rm ep}$ and $B_0$, which are 0.15 and $3\,\mu$G, respectively, for our fiducial model shown in Figure~\ref{fig:radio_flux}. In particular, the magnetic field is determined by the absence of any spectral break up to the highest frequency point detected by \textit{Planck} at 30\,GHz. For our fiducial model the breaking frequency is located at $\sim 200$\,GHz and below such energy the emission behaves like a single power-law. 
Values of $B_0$ much smaller than $3\,\mu$G are incompatible with the radio data as shown in Figure~\ref{fig:radio_flux}, where we also display the synchrotron emission from a model with $B_0=1\,\mu$G, keeping all the other parameters unchanged. Such a low magnetic field implies $\nu_{\rm br} \simeq 60$\,GHz and underestimates the synchrotron emission as measured by \textit{Planck}. In addition $B_0= 1 \,\mu$G implies $K_{\rm ep}=0.85$, a very large value that has never been inferred in other SNRs.

On the contrary, $B_0 \gg 3\mu$G is still compatible with the radio data, but it would violate the pressure equilibrium. In fact, the detection of H$\alpha$ emission from several regions of the forward shock \citep{Blair_2005} implies that the temperature of the CSM has to be $\approx 10^4$\,K, hence the ratio between magnetic and thermal pressures is $P_{\rm mag}/P_{\rm gas} \approx 0.6 \, (B_0/3\mu{\rm G})^2/(n_0/0.4\, {\rm cm}^{-3})$ implying that $B_0$ cannot be much larger than the value adopted here.

In Figure~\ref{fig:mag_field}, we show the profile of magnetic field strength inside and outside the SNR at $t=t_{\rm Sed}$ and at $t=t_{\rm age}$. In the first case, the amplification is very efficient, giving $\delta B_1/B_0 = 6.5$, which after the compression at the shock rises up to $B_2 = \sqrt{11} B_1 \simeq 65\,\mu$G,  while for $r<R_{\rm sh}$ decreases due to adiabatic losses. In addition, a precursor upstream of the shock develops with a thickness $L_{\rm pr} \simeq 0.05 R_{\rm sh}(t_{\rm Sed})$. On the contrary, the amplification is inefficient at the present age, resulting in $\delta B_1/B_0 = 0.06$ and $B_2 \simeq 10\,\mu$G. A precursor is still present with $L_{\rm pr} \simeq 0.08 R_{\rm sh}(t_{\rm age})$ but is not visible in the plotted scale. The peak observed at $r/R_{\rm sh} = 0.73$ corresponds, instead, to the plasma that has been shocked at the $t=t_{\rm Sed}$, namely when the magnetic field amplification was much more efficient. Nevertheless, the contribution of this region to the overall radio emission today is negligible. The same profile is also shown for the case with $B_0= 1\,\mu$G. Interestingly, the peak value at $r/R_{\rm sh} = 0.73$ is very close to the case with $B_0=3\,\mu$G, confirming that the magnetic field there was dominated by the amplification process. 

In the Appendix~\ref{sec:precursor}, we demonstrate how the inferred level of turbulence in the precursor is compatible with that being determined by the competition between the CR amplification and the damping due to ion-neutral friction. At the same time, this damping can explain the temperature of the shock precursor as measured through Balmer emission.

Concerning the energy budget, the total content in the non-thermal particles located inside the SNR is $8.5 \times 10^{48}$\,erg  and $5.7 \times 10^{49}$\,erg for electrons and protons, respectively. The total energy in magnetic field instead is $4.7 \times 10^{47}$\,erg. Hence, it is clear that the equipartition argument illustrated in Sect. \ref{Sec: Cyg Loop spectral analysis} does not hold. This is not surprising, given that a SNR is a transient system where non-thermal particles do not have enough time to equilibrate with the magnetic field which is, instead, a byproduct of instabilities involving mainly ions.

Figure\,\ref{fig:radio_flux} shows that Cygnus Loop is still producing non-thermal X-ray emission thanks to the residual component of non confined electrons. Such flux is much smaller than the detected X-ray emission, estimated to be $56.9 \times 10^{-11} \rm erg \,s^{-1}\, cm^{-2}$ in the  [0.7-1.85] keV interval \citep{Tomida_2016} which is, however, compatible with being of purely thermal origin. In this work we do not attempt to model the thermal emission because it requires to account for the different chemical abundances to fit the X-ray lines, and goes beyond our aims. Nevertheless, as a consistency check, we can compare the electron temperature inferred from X-ray data with the proton temperature estimated from our model. X-ray emission detected with ROSAT from the whole SNR allow to infer an electron temperature of $60_{-20}^{+40}$\,eV \citep{Levenson_1999}. In our model, the post shock proton temperature is $k_B T_p = 3/(16)\, m_p u_{\rm sh}^2 = 280$\,eV, which implies that the electron to proton temperature ratio is $T_e/T_p = 0.2_{-0.07}^{+0.14}$. Such a result is compatible with the finding that for shock speed below $\sim 10^3$\,km s$^{-1}$ the equilibration between electron and proton is quite efficient, resulting in $T_e/T_p > 0.1$ \cite[see, eg.][]{Ghavamian_2001}.

A final comment concerns the value of $K_{\rm ep}$ which is found to be 0.15, quite larger than typical values estimated for young SNRs (like Tycho or SN 1006), which are in the range $10^{-4}-10^{-2}$ \citep{Berezhko+2013ICRC,Morlino-Caprioli2012,Morlino+2009}. This finding may be a peculiarity of the Cygnus Loop or may indicate that the electron/proton ratio is larger for middle-aged SNRs, suggesting that $K_{\rm ep}$ increases for decreasing shock speed. Indeed, the mechanism allowing electrons to be injected into the DSA is still far from being understood. Given the small Larmor radius, thermal electrons need to be pre-accelerated before to enter the DSA. Such pre-acceleration is probably due to kinetic instabilities that develop in the shock foot, as shown by particle-in-cell (PIC) simulations  \citep{Amano-Hoshino2010,Riquelme-Spitkovsky2011}, which suggest that electron injection increases for increasing shock speed \citep{Xu-Spitkovsky-Caprioli2019}, contrary to our suggestion. Nevertheless, at the moment, PIC simulations are limited to explore 1D systems, and cannot catch the complex phenomenology of 3D reality. 
It is also important that $K_{\rm ep}$ may be lower than what estimated here if additional mechanisms able to amplify the magnetic field are present downstream of the shock, like turbulent dynamo processes \citep{Giacalone-Jokipii2007}. In such a case, the downstream magnetic field increases without modifying the proton maximum energy (which is mainly determined by the upstream magnetic field). Hence, the hadronic $\gamma$-ray emission would remain unaltered but the number of electrons required to match the radio data would be reduced because the product $B_2\,n_e$ is fixed by the radio flux.
A systematic study of SNRs showing possible correlation between the electron/proton ratio with SNR properties can shed light on this difficult problem.

\section{Conclusions}
\label{Sec: conclusion}

We presented the spectral analysis performed on the whole Cygnus Loop SNR and on its two peculiar regions, NGC 6992 and the southern shell, by using the Medicina and SRT data between 7.0 and 24.8~GHz and the \textit{Planck} data at 30 and 40 GHz.

Our observations at 8.5 GHz of the entire Cygnus Loop SNR confirm the tendency suggested by the \textit{Planck} data, ruling out any spectral curvature up to high radio frequencies. By modelling the radio and  $\gamma$-ray emission, we constrained the maximum particle energy (for both electrons and protons) at 65~GeV, and the magnetic field strength at the shock at 10~$\mu$G. The model description of the radio data also constrains the electron density, revealing a dominant IC emission from escaping electrons in the $\gamma$-ray spectrum above $\sim$10~GeV. This result sheds new light on the electron contribution to the SNR spectral features at high energies. In this respect, new $\gamma$-ray observations with the next IACT generation, like the Cherenkov Telescope Array\footnote{\hyperlink{https://www.cta-observatory.org}{https://www.cta-observatory.org}} (CTA), could be crucial to investigate the emission produced through IC by escaping electrons at energies $\sim\!1$~TeV and place observing constraints on this spectral tendency. On the other hand, high-frequency radio data are also required to better constrain the electron density and the background magnetic field $B_0$. The new multi-feed $Q$-band (33-50~GHz) receiver \citep{Navarrini_2016}, that is currently being implemented for SRT, will allow us to explore this range.

We investigated the integrated spectrum of NGC 6992 and the southern shell between 7.0 and 44~GHz, by using our single-dish data and the \textit{Planck} flux density measurements.
Both regions present a spectrum flatter than the one associated with the whole Cygnus Loop SNR, showing no indication of a spectral cutoff. The \textit{Planck} data indicated a flux density rising at 44 GHz in the NGC 6992 spectrum that we attributed to a significant dust-emission contribution. In the case of the southern shell, \textit{Planck} measurements suggested a concave-up spectral shape that we tentatively ascribed to an efficient production of CRs, which could modify the shock structure. For both regions, further sensitive flux density measurements in the range between $\sim8$ and $\sim50$~GHz are needed to firmly characterise their spectral tendency.

Unlike the case of the whole Cygnus Loop SNR, we could not model the non-thermal emission from NGC 6992 and the southern shell, because no separated $\gamma$-ray measurements are available for these regions. Nevertheless, we exploited the fact that the radio spectrum of both regions showed no indications for a spectral steepening to establish a lower limit on the break frequency of $\sim$~25 GHz and estimate the magnetic field strength under the assumption of equipartition between the particle and magnetic field energy. For the southern shell, we obtained a value very similar to  that estimated in the same way for the whole SNR, suggesting that this region is a good approximation for the properties of the whole SNR.
In the case of NGC 6992, we obtained a higher magnetic field with respect to the average Cygnus Loop value. In addition, this region presents a harder radio spectrum with respect to the whole SNR. Both these findings may be a consequence of the fact that the shock is currently transiting to a radiative phase. However, these hypotheses can be tested only through a proper modelling of the non-thermal emission associated with this region possibly accounting for the presence of thermal Hydrogen which can significantly modify the shock dynamics \citep{Morlino+neutrals:2013}. In this regard, high spatial resolution $\gamma$-ray observations, which will be possible with CTA, will be crucial to achieve spatially-resolved images of the Cygnus Loop and constrain the maximum particle energy and magnetic field strength of its peculiar regions through refined models, like the one used in this work for the whole SNR.

\section*{Acknowledgements}

The Sardinia Radio Telescope is funded by the Department of University and Research (MIUR), the Italian Space
Agency (ASI), and the Autonomous Region of Sardinia
(RAS), and is operated as a National Facility by the National Institute for Astrophysics (INAF). 
We deeply thank W. Reich for assistance and the very useful suggestions.
We are very grateful to H. Katagiri for the useful discussion about the \textit{Fermi}-LAT data. SL acknowledge contribution from the grant INAF CTA-SKA "Probing particle acceleration and $\gamma$-ray propagation with CTA and its precursors".
GM acknowledges support from Grants ASI/INAF n. 2017-14-H.O, SKA- CTA-INAF 2016 and INAF-Mainstream 2018.


\section*{Data availability}

The data underlying this article are all reported in the tables of the paper.



\bibliographystyle{mnras}
\bibliography{mnras_template.bib} 




\appendix

\section{The shock precursor}
\label{sec:precursor}
One of the main finding of our acceleration model is the fact that the particle  maximum energy in the ST stage decreases in time like $\sim t^{-3}$. This is quite remarkable because the streaming instability (resonant and non resonant) predicts a less pronounced decrease, with $\delta \leqslant 2$. One possible explanation for our finding is connected to the presence of neutral Hydrogen in the CSM, which in the case of Cygnus Loop is firmly established by the detection of H$\alpha$ emission \cite[see, e.g.][]{Blair_2005}. In fact, neutral atoms can damp the amplified magnetic waves, making the particle escape  from the source easier. Here, we show that this assumption is compatible with the level of magnetic turbulence that we have found and above which, quite remarkably, can also account for the temperature increase observed in the precursor ahead of the shock \citep{Katsuda_2016}.

We start from the common assumption that the magnetic turbulence upstream consists of Alfv\'en waves excited by streaming instability with a rate (integrated over all momenta) given by \citep{Skilling1975}
\begin{equation}
    \Gamma_{\rm cr}=  \frac{4 \pi v_A}{\delta B^2} \frac{\partial P_{\rm cr}}{\partial x}\,,
\end{equation}
where $v_A= B_0/\sqrt{4\pi \rho_i}$ is the Alfv\'en speed, $\rho_i$ is the ion mass density and $P_{\rm cr}$ the CR pressure at the shock.
At the same time, the turbulence is damped by the ion-neutral friction with a damping rate \citep{Kulsrud-Cesarsky1971}:
\begin{equation}
    \Gamma_{\rm in} = \frac{\nu_{\rm in}}{2} 
    = 4.2 \times 10^{-9} \frac{n_n}{\rm cm^{-3
    }} \left(\frac{T}{10^4 \rm K}\right)^{0.4} \; \rm s^{-1} \,,
\end{equation}
where $\nu_{\rm in} = n_n \langle \sigma v \rangle$ is the ion neutral collision frequency, $n_n$ is the neutral Hydrogen density and $T$ the plasma temperature. For reasonable values of $n_n$ and $T$, the damping timescale $\tau_{\rm in} = \Gamma_{\rm in}^{-1}$ ranges between few years to few tens of years, hence it is much smaller than the advection time in the precursor, given by $t_{\rm adv} = D_1(p_{\max})/u_{\rm sh}^2 \simeq 2600$\,yr. In such a situation, the level of magnetic turbulence is given by the equilibrium condition between damping and growth, i.e.  $\Gamma_{\rm in} = \Gamma_{\rm cr}$. Approximating $\partial_x P_{\rm cr} \approx P_{\rm cr}/L_{\rm pr}$, where $L_{\rm pr}=D_1(p_{\max})/u_{\rm sh}\simeq 1$\,pc is the CR precursor length, the equilibrium condition gives:
\begin{equation}
    \left(\frac{\delta B_1}{B_0} \right)^2 
    = \frac{P_{\rm cr}}{B_0^2/(4\pi)} \, \frac{v_A}{ u_{\rm sh}} \, \frac{\tau_{\rm in}}{t_{\rm adv}} \,.
\end{equation}
Using the parameter values in Table~\ref{tab:parameters} and assuming a neutral fraction of 0.5, we found $\delta B_1/B_0=0.18$, namely 3 times larger than the value found in Sect. \ref{sec:mod_results}. Given the uncertainties in several parameters, we consider this result in fair agreement with our model.

An interesting method to test the level of magnetic turbulence in the shock precursor is through the measurement of its temperature. 
\cite{Blair_2005} and \cite{Katsuda_2016} detected Balmer emission from non-radiative shocks in the northeastern part of the Cygnus Loop, reporting a clear detection also in the region upstream of the shock. Measuring the H$\alpha$ line width, they inferred a  temperature of $\sim 18,000$\,K in an extended region of $\sim 0.1(d/735 {\rm pc})$\,pc attributed to a photo-ionization precursor plus a smaller region close to the shock with $T \sim 35,000$\,K and size $\sim 5.5\times 10^{15} (d/735 {\rm pc})$\,cm, which they attribute to a tiny CR precursor. However, temperatures $T > 15,000$\,K lead to complete ionization of neutral hydrogen in equilibrium, so that H$\alpha$ emission from the filament would not be detected. Thus, the reported observation implies that the pre-shock gas is heated in a thin precursor. 
\cite{Katsuda_2016} argued that the damping of Alfv\'en waves could account for at least a fraction of this heating, and here we want to investigate this scenario. The total energy density released by the turbulent damping into a plasma element that crosses the whole  precursor is $\Delta U = \delta B^2/(8\pi) \, \Gamma_{\rm in} t_{\rm adv}$. This energy is completely converted into heating, hence, neglecting radiation losses, the final plasma temperature just ahead of the shock is given by the relation $\Delta U = 1.5 n_0 k_B (T_{\rm pr}-T_0)$, where $k_B$ is the Boltzmann constant and $T_0\approx 10^4$\,K is the far upstream temperature. Using the same values as above, we have
\begin{equation}
    T_{\rm pr} = T_0 + \frac{3 \, \xi_{\rm cr} \, u_{\rm sh} \, v_A}{k_B}
    \simeq T_0 + 1.6 \times 10^4 \,{\rm K}
\end{equation}
a value in between the extended and the small precursors' temperatures. This is a quite remarkable result, in that we have not tuned any parameter of the model to get it, and demonstrates that Alfv\'en wave damping can be the main process to heat the shock precursor.
One should note, however, that the CR precursor length that we estimated is $\sim 10$ times larger than the extended precursor and $\sim 500$ times larger than the smaller and hotter precursor.
Hence the extended precursor could be mainly heated by CRs rather than by ionizing photons. On the other hand, the thinner precursor requires an additional heating which may be provided either by very low energy CRs ($E\lesssim 30$\,MeV) or by fast neutrals \citep{Katsuda_2016}. A proper answer can be provided by using a model for particle acceleration in presence of neutral plasma \citep{Morlino+neutrals:2013}.


\bsp	
\label{lastpage}
\end{document}